\documentclass[1p]{elsarticle}

\usepackage{amssymb}
\usepackage{amsmath}
\usepackage{amsfonts}
\usepackage{graphicx}
\usepackage{epstopdf}
\usepackage{epsfig}
\usepackage{lineno}
\usepackage{xcolor}
\usepackage[normalem]{ulem} 
\usepackage{url}

\newcommand{\Eloc}{h}

\def\cn2{|c_n|^2}

\newcommand{\ii}{\textrm{i}}

\renewcommand{\d}{\textrm{d}}

\begin{document}


\begin{frontmatter}
\title{Thermal lifetime of breathers}%

\author[gfnl]{Juan F. R. Archilla\corref{cor1}}
 \ead{archilla@us.es}
\cortext[cor1]{Corresponding author}
 \address[gfnl]{Group of Nonlinear Physics, Universidad de Sevilla,\\ ETSII, Avda Reina Mercedes s/n, 41012-Sevilla, Spain}
\author[lvu]{J\={a}nis Baj\={a}rs}
 \ead{janis.bajars@lu.lv}
 \address[lvu]{Faculty of Science and Technology, University of Latvia,\\
 Jelgavas Street 3, Riga, LV-1004, Latvia}
\author[ibs]{Sergej Flach}
 \ead{sflach@ibs.re.kr}
 \address[ibs]{Center for Theoretical Physics of Complex Systems, Institute of Basic Science,\\ Expo-ro 55 Yuseong-gu,
Daejeon 34126, Republic of Korea}

\begin{abstract}
In this article, we explore the lifetime of localized excitations in nonlinear lattices, called breathers, when a thermalized lattice is perturbed with localized energy delivered to a single site. We develop a method to measure the time it takes for the system to approach equilibrium based on a single scalar quantity, the participation number, and deduce the value corresponding to thermal equilibrium.  We observe the time to achieve thermalization as a function of the energy of the excited site. We explore a variety of different physical system models. The result is that the lifetime of breathers increases exponentially with the breather energy for all the systems. This increase becomes observable when this energy is larger than approximately ten times the local average thermal energy. These results may provide a method to detect the existence of breathers in real systems.
\end{abstract}
\begin{keyword}
nonlinear waves \sep breathers \sep thermal equilibrium \sep localization
\PACS  63.20.Pw 
 \sep 63.20.Ry  
\sep 05.45.-a	
02.70.-c 
\end{keyword}
\end{frontmatter}

\newcommand{\fracc}[2]{\frac{\displaystyle #1}{\displaystyle #2}}


\section{Introduction}
Discrete breathers (DB) are exact localized vibrations in nonlinear lattices\,\cite{mackayaubry94,flach1998,2004PhT....57a..43C,flach2008}. They are also called {\em Intrinsic Localized Modes} (ILM)\,\cite{sievers-takeno1988}, to distinguish them from the localized Anderson modes due to disorder, an external impurity, a defect, or an interface in a lattice\,\cite{anderson1958, archilla1999}. It is important to note that this definition applies only in the absence of perturbations, noise, or temperature, conditions for which only approximate breathers are possible. 

Usually, discrete breather solutions cannot be obtained in closed analytical form. However, they can be constructed numerically with arbitrary machine precision, and, in that case, they are called {\em exact breathers}\,\cite{flach1995,marinaubry96,marin1998,archilla2001}. Numerical simulations of the evolution of exact solutions can continue forever at zero temperature. This is because exact DB solutions are usually exponentially localized on an infinitely large lattice, such that the energy of these solutions is finite, and their average energy density is exactly zero. However, introducing the concept of temperature or simply finite energy density usually implies a finite lifetime for any coherent excitations, which is an important feature to measure or estimate when studying breathers in physical systems.

The existence and lifetime of breathers can be of importance in environments where they will be created in large quantities, as fusion tokamak reactors\,\cite{itertokamak2023}, where the impact of neutrons and alpha particles will produce many of them. Their slow relaxation may prevent the evacuation of heat at the desired rate with undesirable consequences. The effect could be even more remarkable if breathers can bind to an electric charge in the materials used\,\cite{russell-archilla2024}.

An early attempt to quantify the lifetimes of DBs on a nonzero thermal background was reported in \cite{IvanchenkoDiscretePhysicaD2004}, where simulations were performed to detect long-lasting local fluctuations and their lifetimes as a function of the energy density of the background. The fluctuation lifetime grew with the energy density, seemingly in an exponentially fast way. This approach was clearly not capable of measuring the lifetimes of strongly excited breathers due to computational limitations. Also, the energies of the longest-lasting local fluctuations depended on the computational time - the longer the time, the more probable a larger fluctuation becomes.

Molecular dynamics was used in different publications to measure the lifetime of gap breathers in thermal equilibrium. Ref.\,\cite{khadeeva-dmitriev2011} considers a two-dimensional diatomic crystal. Wavelet imaging was used for nonlinear excitations that appear in the phonon gap of a NaI crystal\,\cite{riviere-piazza2019}. In both systems, the lifetime appears to increase with temperature, which is consistent  with the creation of breathers with larger energies.
The approach in the present work is different from the previous works since we consider the initial localization of energy {\em not} at thermal equilibrium.

A  recent study by Iubini et al \cite{iubini2019} planted local breather excitations into a discrete nonlinear Schr\"odinger equation (DNLS). They were motivated by the fact that the microcanonical DNLS dynamics features thermal states with Gibbs statistics but also non-Gibbs ones with nonergodic features \cite{rasmussen00,rumpf2008,rumpf09,PhysRevLett.120.184101,ChernyNOn-GibbsPhysicalReviewA2019} due to the presence of an additional (to the already existing energy) integral of motion related to the total norm or a classical analog of the total number of particles of quantum many-body systems. The study monitored the local norm at the original site of the breather excitation, introduced an ad hoc threshold to its value at which the breather was assumed to be destroyed, and measured the time to reach that threshold. The study found approximate exponential dependence of the breather's lifetime on its norm.

Finally, around the same time, Danieli et al \cite{DanieliCampbellFlachPRE2017} introduced a method to define and measure the lifetime of the fluctuation of any observable using the ergodic hypothesis. They applied this method successfully to measure the lifetime of long wavelength mode excitations in Fermi-Pasta-Ulam-Tsingou (FPUT) systems to address the FPUT paradox of lifetimes of anomalous initial states and quenches, but also to address the statistics of lifetime fluctuations at thermal equilibrium. We are going to use this method below.

In this article, we address the issue of breather lifetimes by studying systems that are more general than the DNLS one, which in general lack nontrivial second integrals of motion, and which are supposed to show proper Gibbs statistics and ergodicity for any choice of the relevant energy density. Instead of an ad hoc threshold value for local energy, we use the participation number $P$ as an observable and measure the inhomogeneity of energy density distribution in the system. At thermal equilibrium, the participation number $P$ on any ergodic trajectory is forced to fluctuate around its temporal and thus phase space average endlessly. We compute the thermal average of $P$ and measure the time interval it takes for a breather excitation on top of a thermal background to decrease the participation number down to its thermal average for the first time. We find strong exponential dependence of the thermalization time as a function of the breather excitation energy by studying four different, fairly generic systems, consisting of units described by a coordinate $u_n$, which can be a spacial coordinate or a different magnitude, and its momentum $p_n$. For simplicity, we will often refer to each unit as an oscillator or particle. Moreover, we find that the exponential dependence is strongly influenced by the temperature, i.e. energy density, of the system background.

The first three systems represent a lattice of particles that experience an on-site potential $V(u_n)$, where $u_n$ is the coordinate of the $n^{th}$ particle, and the particles are coupled with their nearest neighbors through an interaction potential $U(u_{n+1}-u_n)$. The on-site potential represents an external field, or, if the described system is a subsystem of a larger system, the interaction with the rest of the system. The first two systems have a quartic on-site potential and harmonic coupling, that is:
\begin{equation}
H=\sum_n \frac{1}{2}p_n^2+ V(u_n)+U(u_{n+1}-u_n)=
\sum_n \fracc{1}{2}p_n^2+ \omega_0^2\left(\frac{u_n^2}{2}+s \frac{u_n^4}{4}\right)
+\kappa \fracc{1}{2}(u_{n+1}-u_n)^2\,,\label{eq:quartic}
\end{equation}
where the nonlinearity parameter  $s$ can be $+1$ or $-1$. If $s$ is positive, an isolated particle will have a frequency that increases with the oscillation amplitude, or, as it is commonly called, a {\em hard} potential. This system, labeled QH, will be the first system studied, followed by the opposite case, that is, the quartic {\em soft} potential (QS) with $s=-1$, which becomes the second system. The third, more realistic system, has a Frenkel-Kontorova on-site potential\,\cite{braun1998, braun2004}, that is, a sinusoidal function, which models the periodicity of the lattice. The coupling potential is the Lennard-Jones potential, which represents a realistic interaction between atoms, that repel very strongly when they get close and gradually vanishes when they move apart\,\cite{lennard-jones1925, lennard-jones1929,schwerdtfeger2024}. We will denote it as FKLJ. The fourth system will have no on-site potential, but a sinusoidal type coupling, and represents a Josephson junction network\,\cite{barone1982,lando-flach2023}. As the coordinate describing an oscillator is an angle, it is also called a {\em rotor}, and the system can also represent a chain of coupled pendula\,\cite{barone1982} that can rotate. We will often refer to this analog as it is easier to understand. We will denote this system as JJN.

In this study, the first step is to obtain information about the discrete breather energies and characteristics. In the appendices, we describe the method to obtain {\em exact} breathers from the anticontinuous limit and their properties in the four different systems.

The paper is organized as follows: in Sec.\,\ref{sec:thermalization}, we introduce the participation number $P$, deduce its value at thermal equilibrium, and demonstrate that numerical simulations are coherent with this interpretation. Sec.\,\ref{sec:evolution} describes the creation of localized breather energy over a thermalized system and the system's time evolution towards thermal equilibrium. The following Section \ref{sec:lifetime} is dedicated to the computation of breathers' lifetime and analysis of differences in numerical results in the systems under study. The different systems are analyzed in four subsections: the quartic hard and soft potentials in Sec.\,\ref{sec:qh} and Sec.\,\ref{sec:qs}, respectively, the Frenkel-Kontorova Lennard-Jones system in Sec.\,\ref{sec:fklj} and the model for Josephson junction arrays in Sec.\,\ref{sec:jjn}. The paper concludes with the conclusions, acknowledgments, funding, and a short reference to the computational means that have been used. The appendices include analytical and computational details about discrete breathers in the four different systems.
\begin{figure}[ht]
\begin{center}
\includegraphics[width=0.49\textwidth]{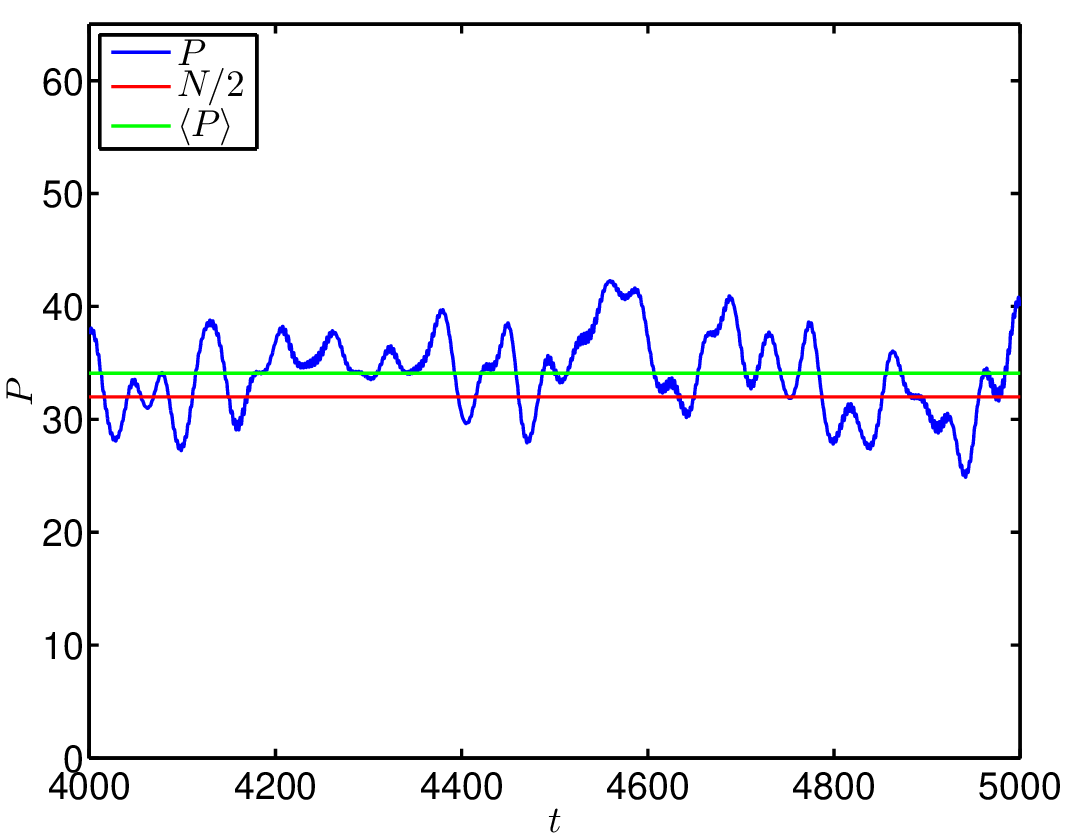} 
\includegraphics[width=0.49\textwidth]{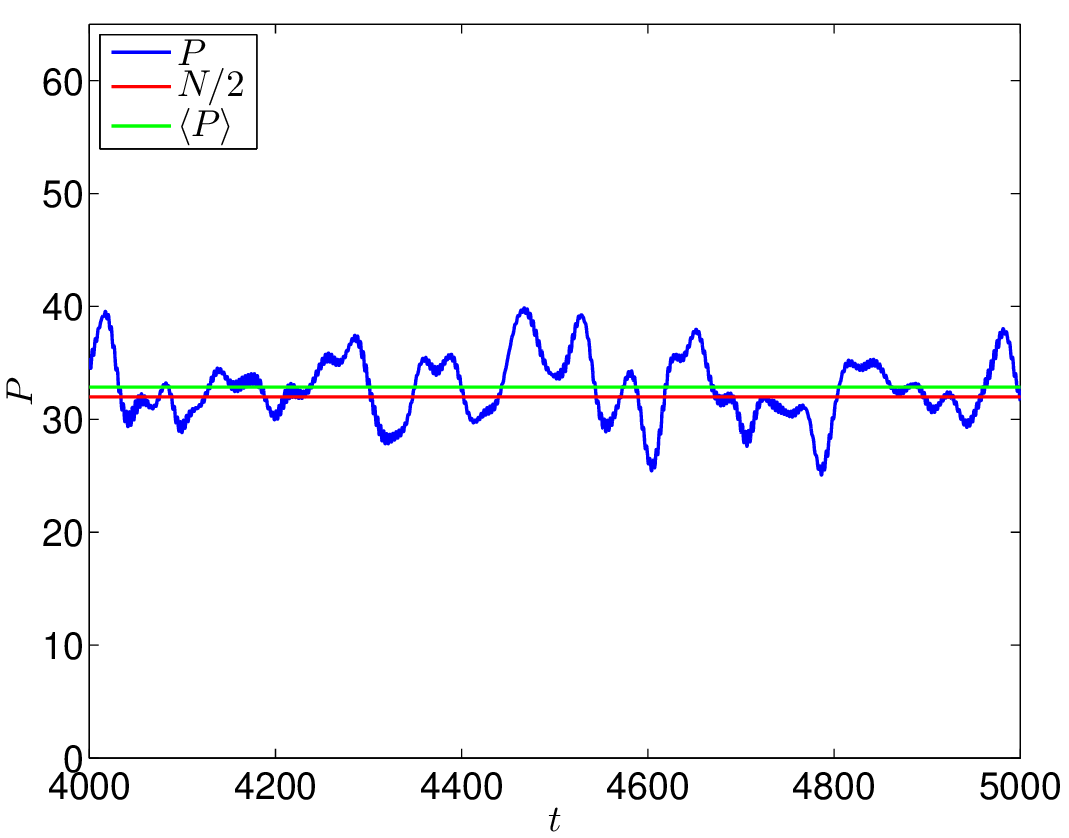} 
\end{center}
\caption{ ({\bf Left:}) System QH with $\omega_0=1$ and $\kappa=0.05$: Evolution of the participation number $P$ for the mean thermal energy $\Eloc=0.02$ with the hard quartic potential and with number of particles $N=64$. We can see that the time-averaged participation number $\langle P\rangle$ is slightly above $N/2$.
({\bf Right:}) System QS: Identical representation for the soft potential with the same parameters. In both cases, generally speaking,  $\langle P\rangle$ is 1-2 particles above $N/2$, but for some realizations, it can also be below depending on the length of the observation time window.
}
\label{fig_thermal02}
\end{figure}

\section{Description of thermalization}
\label{sec:thermalization}

We present here a suitable parameter to measure the thermalization state of the system and explain the procedure to study the thermalization of breathers and their lifetime. The considerations in this section are generic, but we will illustrate them mainly with the QH and QS models of Eq.\,\eqref{eq:quartic}, with $s=\pm 1$.

\begin{figure}[ht]
\begin{center}
\includegraphics[width=0.48\textwidth]{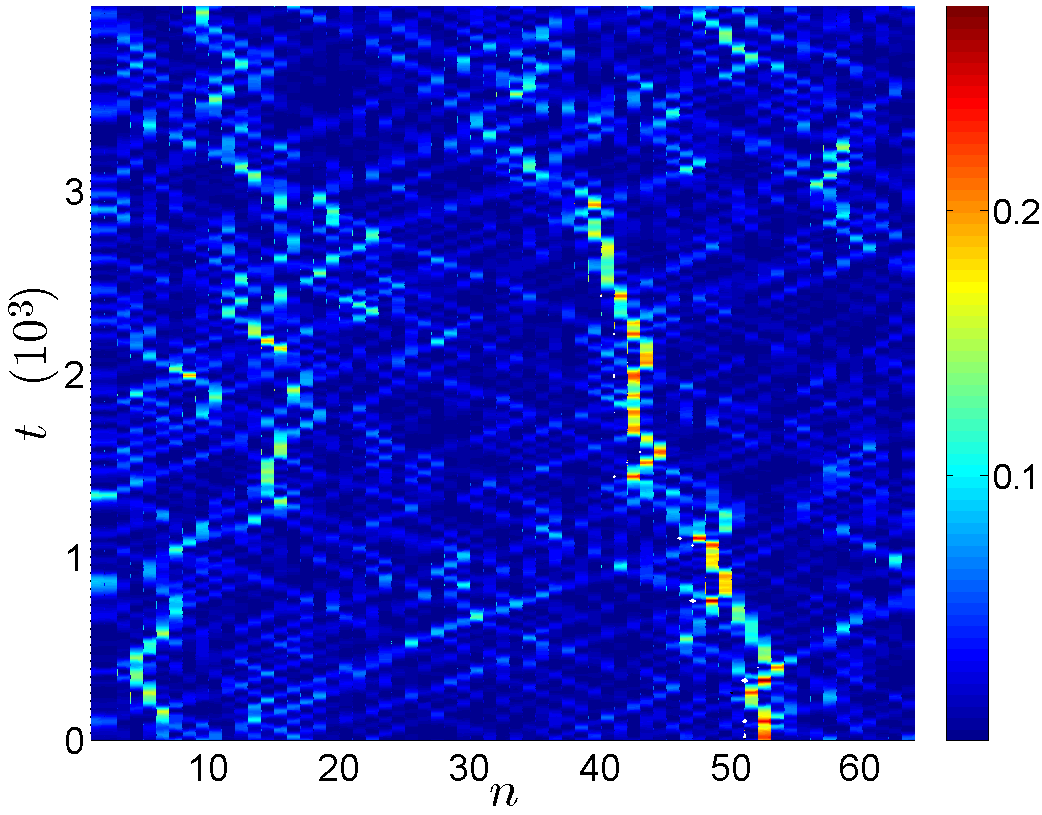}\, 
\includegraphics[width=0.48\textwidth]{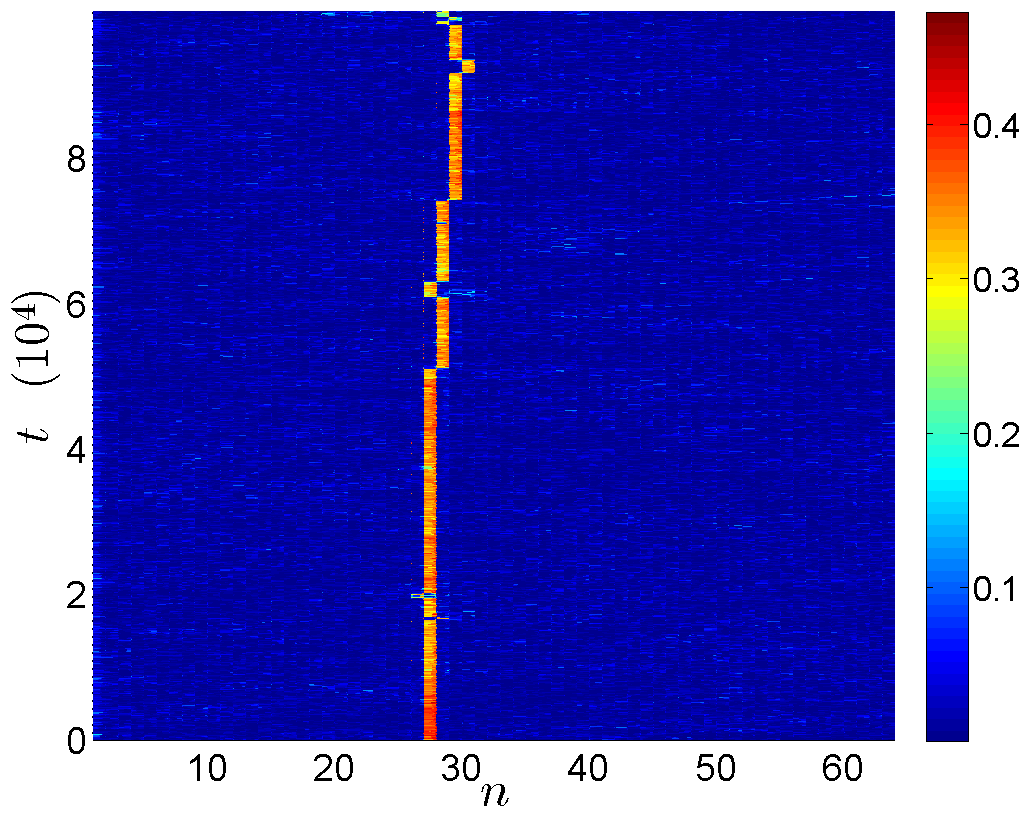} 
\\
\includegraphics[width=0.48\textwidth]{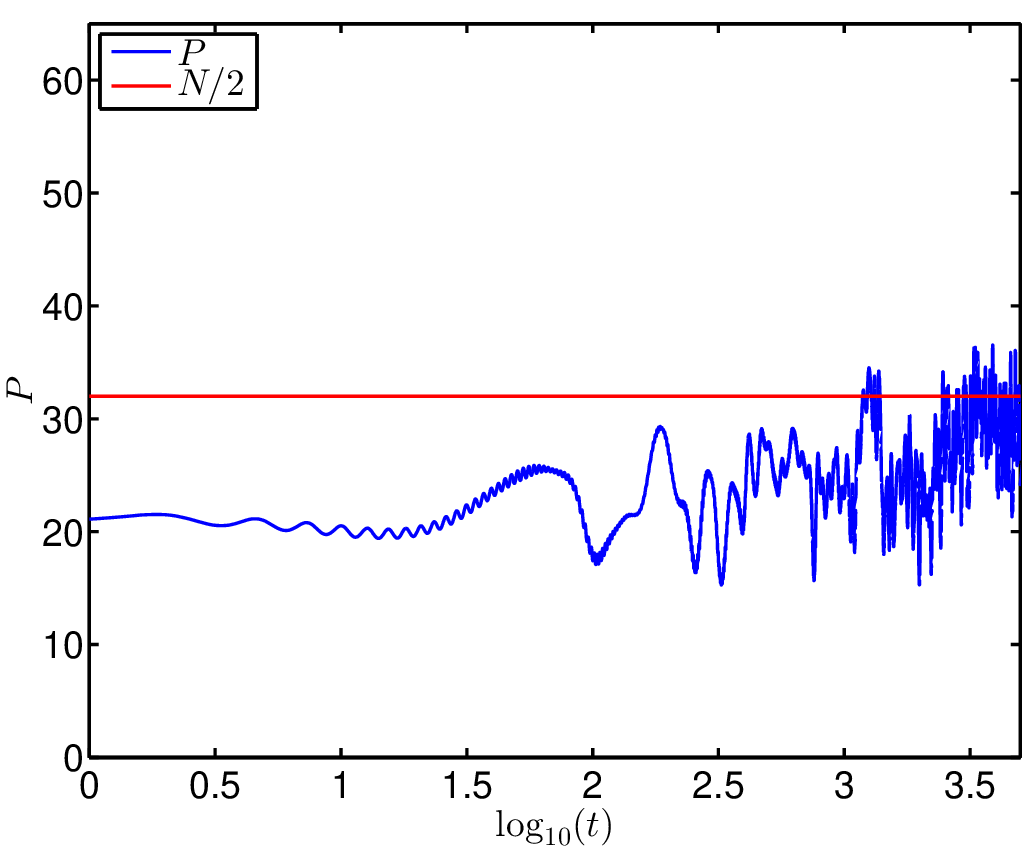}\,
\includegraphics[width=0.48\textwidth]{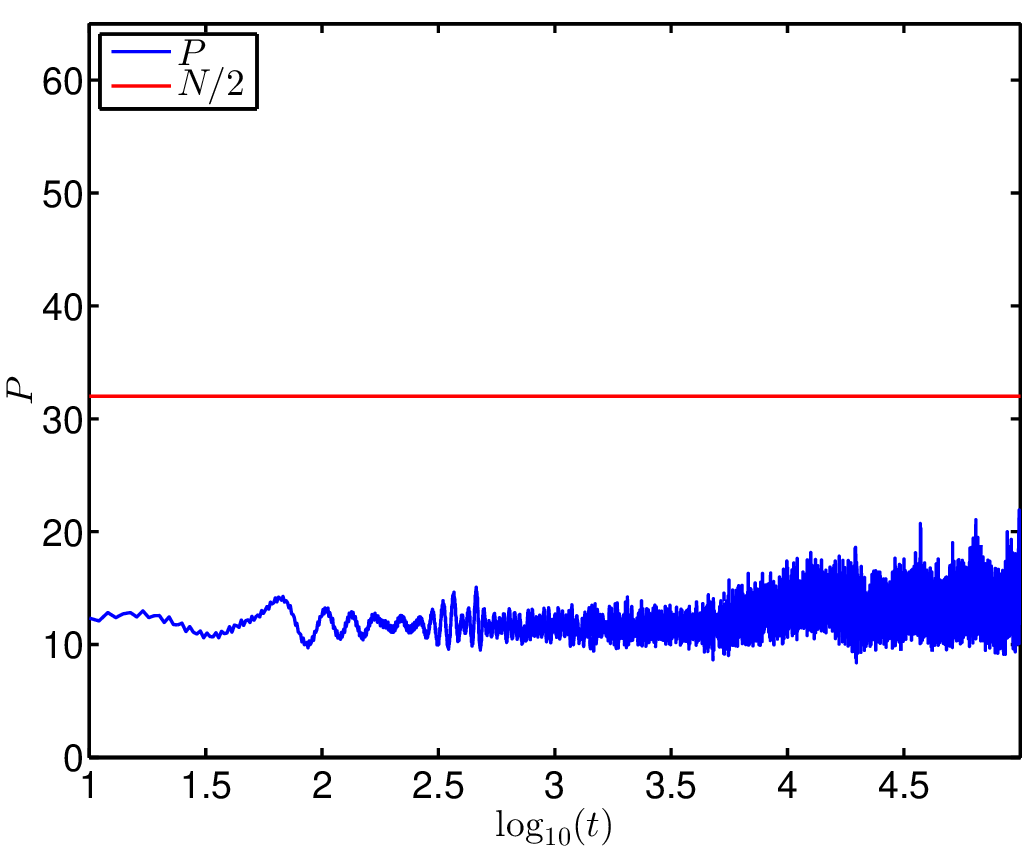} 
\end{center}
\caption{  ({\bf Top left:})
System QH with $\omega_0=1$ and $\kappa=0.05$: Evolution towards equilibrium of a breather in a thermalized system with $N=64$ particles, $\Eloc=0.02$ and $E_\text{b}=10\Eloc$. The thermalization is achieved relatively soon. ({\bf Top~right:})~System QH with the same parameters except the breather energy
 $E_\text{b}=21 \Eloc$. The breather is kept well localized although it hops positions but without the system achieving equilibrium during the observation time. Note the different time span. ({\bf Bottom: left-right}) Corresponding evolution of the participation number $P$. The  operational definition of the thermalization time $t_{th}$ is the first time when $P(t) = N/2$.
 }
\label{fig_brthermal1}
\end{figure}

\subsection{Participation number and thermal equilibrium}
We are interested in a simple magnitude that approximately measures the system's arrival to thermal equilibrium. A good candidate for this measure is the participation number defined as
\begin{equation}
P=\frac{\left(\sum e_n\right)^2}{\sum e_n^2}=\frac{E^2}{\sum e_n^2}\, ,
\label{eq:partition}
\end{equation}
where $e_n$ the energy of the $n^{th}$ oscillator,  $E=\sum e_n$ is the total energy of the system, that is, e.g., the Hamiltonian in Eq.\,\eqref{eq:quartic}, and the summation is over all the oscillators, with periodic boundary conditions. The participation number $P$ takes values between $P=1$, when all the energy is concentrated in a single oscillator, and $P=N$, if it is evenly distributed among all the oscillators.

The system has a constant energy, and the statistical ensemble is, therefore, microcanonical. If the system is ergodic and large enough, we may expect to observe canonical distributions for local observables, characterized by some temperature $T$ and constant thermodynamic $\beta=1/k_B T$. Note that although the values of $k_B$ and $T$ depend on the specific units, $k_B T=1/\beta$ is well defined, as the statistical meaning of temperature is given by the virial theorem\,\cite{reif2009} and the average kinetic local energy at thermal equilibrium is $\langle 0.5 p_n^2\rangle=0.5/\beta=0.5 k_B T$.

Each particle can be seen as interacting with a thermal reservoir at constant temperature and the statistical ensemble of each particle is canonical. Therefore, the probability that the energy of the $n^{th}$ particle in thermal equilibrium is $e_n$ is given by\,\cite{reif2009}:
\begin{equation}
\rho(e_n)=\fracc{\exp(-\beta e_n)}{Z}.
\label{eq:rho_en}
\end{equation}
The partition function $Z$ can be obtained with the condition that the total probability of finding some energy within the canonical ensemble is the unity. Then:
\begin{equation}
Z=\int_{0}^{\infty}\exp(-\beta e_n) \, \d e_n=\fracc{1}{\beta}\,.
\label{eq:Zen}
\end{equation}

The actual upper limit of the integral cannot be infinity but a number of the order of the total system's energy $E\sim N k_B T =N/\beta$, with $N$ the number of particles. However, the integral  $\int_{E}^{\infty}\exp(-\beta e_n) \, \d e_n=\exp(-\beta E)\sim \exp(-N)$ becomes negligible as it is smaller than $10^{-10}$ for $N>21$. Thus, \eqref{eq:Zen} and the derivations below should be justified good approximations for the systems with a large number of particles $N$.

The average energy of a particle becomes
$$\langle e_n \rangle =\fracc{1}{Z}\int_{0}^{\infty}\exp(-\beta e_n) e_n \,\d e_n=
\fracc{1}{Z}\fracc{\partial }{\partial \beta}(-Z)=\beta\fracc{\partial }{\partial \beta}\left(-\fracc{1}{\beta}\right)=\fracc{1}{\beta}=k_B T\, ,
$$
as should be expected. This result is obtained with several approximations and the time averages of the local energies at thermal equilibrium differ slightly from it.

In the following, we will use the ergodic theorem\,\cite{reif2009,schwabl2006} and replace the averages within the statistical ensemble with the time averages at thermal equilibrium. This is valid only for infinite time interval averages and, of course, our simulations may be long but always finite. In addition, our approach implicitly assumes that any energy density distribution $\{e_n\}$ has the same probability to occur. This subtle assumption does not follow, in general, from the fundamental ergodicity property that each microstate (some point in phase space) has the same probability of occurring. Indeed, one can easily show that ergodicity results in equal probabilities of energy density distributions for small ratios of coupling to energy density $\kappa/h$ in (\ref{eq:quartic}) and (\ref{eq:Hjjn}). In general, however, we have to expect systematic differences and will check that these differences are smaller than the standard deviations of the temporal fluctuations.

Let us calculate the infinity time average of the inverse of the participation number $P$, that is:
\[
P^{-1} = \frac{\sum_n e_n^2}{E^2}\, .
\]
The time average along any solution of the dynamical system is defined and given in the following form:
\begin{align*}
\overline{P^{-1}} &= \lim_{T\to\infty} \fracc{1}{T} \int_{0}^{T} P^{-1}(s) \, \d s = \lim_{T\to\infty} \frac{1}{T} \int_{0}^{T} \frac{\sum_n e_n^2(s)}{E^2} \, \d s\\
 &= \fracc{1}{E^2}\sum_n \lim_{T\to\infty} \fracc{1}{T} \int_{0}^{T} e_n^2(s) \, \d s =
 \frac{1}{E^2}\sum_n \overline{e_n^2}.
\end{align*}
Through the ergodic theorem, we now identify $\overline{P^{-1}}=\langle P^{-1} \rangle $ and
 $\overline{e_n^2}=\langle e_n^2 \rangle$. Then:
\begin{align*}
\langle P^{-1}\rangle = \fracc{1}{E^2} \sum_n \langle e_n^2 \rangle =\fracc{N (2/\beta^2)}{(N/\beta)^2} = \fracc{2}{N}.
\end{align*}
Numerically we can suppose and it is confirmed by the numerical simulations that
$\overline{P^{-1}} \simeq (\overline{P})^{-1}$ and, therefore, $\langle P^{-1}\rangle \simeq \langle {P}\rangle ^{-1}$,
i.e.,
\begin{equation}
\overline{P}\simeq \langle P\rangle \simeq\fracc{N}{2}.
\label{eq:partitionN2}
\end{equation}
In the rest of the paper we will identify often the average over a long time with the infinitely time averages and the statistical ensemble averages.

Note that the deduction cannot be done directly for $\langle P \rangle$ from \eqref{eq:partition} as the sum of particle energies would appear in the denominator and thus we can not express $P$ as a sum of $\langle e_n^2\rangle$ terms. We have used many approximations that will be ultimately confirmed by numerical simulations. It is interesting to note that the result \eqref{eq:partitionN2} contradicts the first intuition that at thermal equilibrium $P$ would be close to $N$.

It works surprisingly well for the first three systems under study. It does not work so well for the JJN system, because as this system has no on-site potential, the approximation that we can identify the energy of an oscillator as relatively independent is not really valid. See Sec.\,\ref{sec:fklj} for details and Ref.\,\cite{lando-flach2023} for another approach including different approximations.

\subsection{Participation number at thermal equilibrium in simulations}
\label{sec:thermal_eq}

We produce a random vector of momenta $p_n$ values of numbers between $(0,1)$ and subtract its mean $p_n=p_n-\langle \{p_n\}\rangle$ so that the system's initial momentum is zero. We re-scale the momenta so that the mean kinetic energy is the desired mean thermal energy $\Eloc=E/N$, and set the initial coordinates at zero $u=[u_1,\dots,u_N]=[0,\dots,0]$.  After about 100$T_0$ time units, with $T_0=2\pi/\omega_0$, we consider the system thermalized. We let it evolve even for a longer time and compare the mean value of $P$ with the theoretical one $N/2$. For both the hard and soft potentials, $\langle P \rangle$ is generally about one or two particles above the theoretical position of $N/2$ but can also be below it for some realizations depending on the observation time window. This property does not depend significantly on the local energy $\Eloc$. Particular realizations can be seen in Fig.\,\ref{fig_thermal02} for a lattice with $N=64$ particles and systems QH and QS.

We also obtain similar results for the FKLJ system, but for the JJN system, $\langle P\rangle$ is about 6 particles above $N/2$. However, in this case, the fluctuations of $P$ at thermal equilibrium surpass $N/2$ frequently. Being conscious that $N/2$ is, therefore, not such a good measure of thermal equilibrium for the JJN system, we still keep it as a useful measure of the proximity to equilibrium for comparison of the breather thermalization times.

We conclude that the approximate value of $P=N/2$ is an appropriate measure to indicate that the system has approached equilibrium. However, it does not guarantee it, as should not be expected of a single parameter.

\begin{figure}[tb]
\begin{center}
\includegraphics[width=0.485\textwidth]{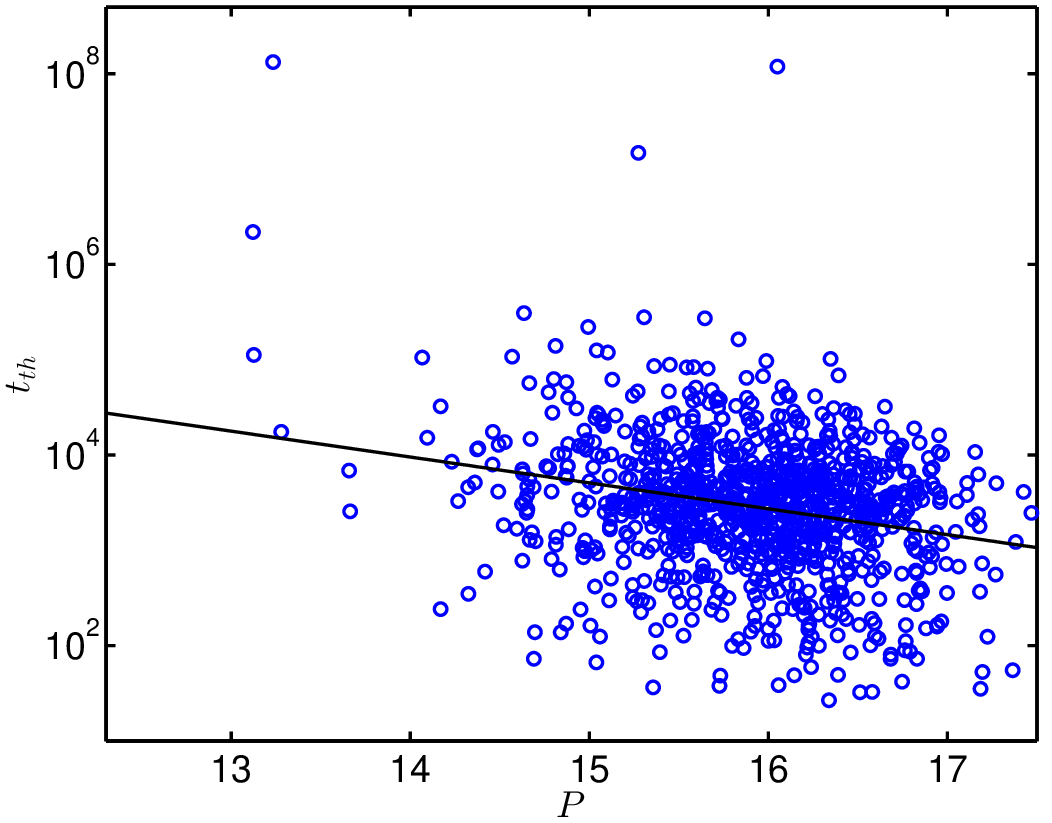}\,\,
\includegraphics[width=0.485\textwidth]{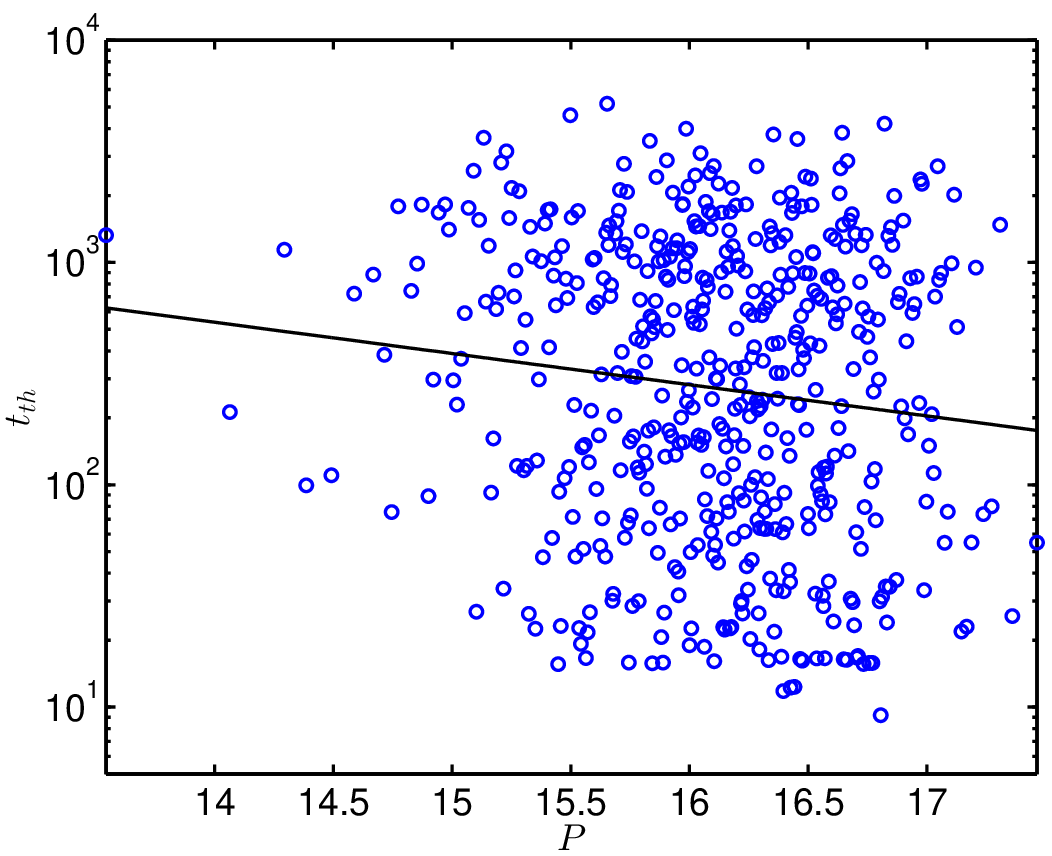}\\
\includegraphics[width=0.485\textwidth]{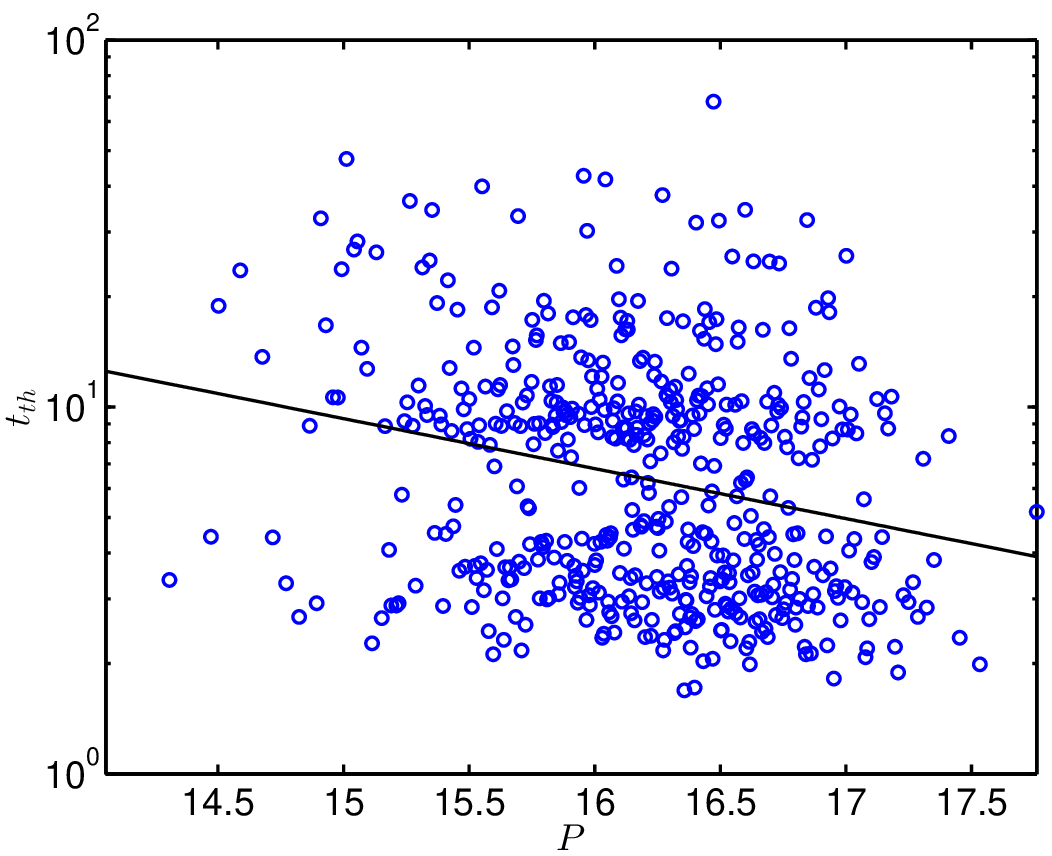}\,\,
\includegraphics[width=0.485\textwidth]{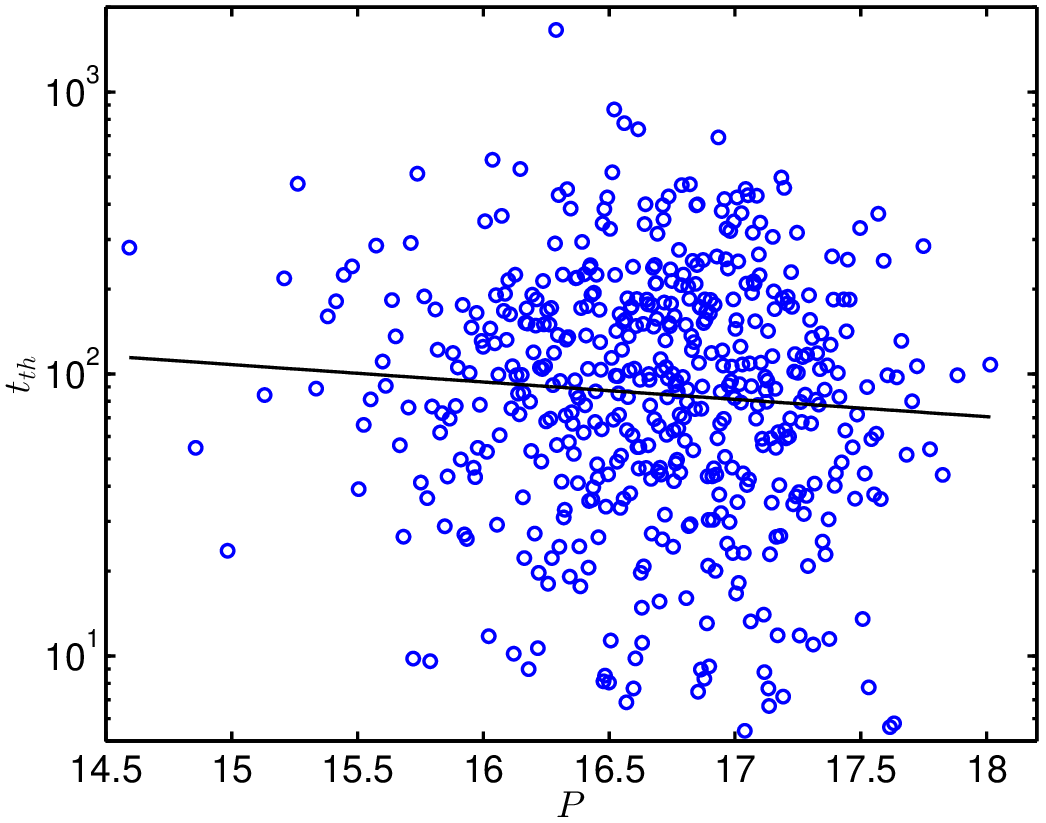}
\end{center}
\caption{({\bf Top-left:}) System QH: Thermalization  time as a function of the {\em initial} participation number $P$ after breather creation, with $N=64$, $\omega_0=1$, $\kappa=0.05$, $\Eloc=0.02$ and $E_\text{b}/\Eloc=16$.
({\bf Top-right:}) System QS: Same representation with identical parameters.
({\bf Bottom-left:}) System FKLJ: Same representation with identical parameters except $\kappa=0.20$.
({\bf Bottom-right:}) System JJN: Same representation with the same parameters as QH and QS. Also, the linear regression line is plotted. Note the different time scales. A random subsample of 500 points is represented for clarity.
 }
\label{fig_Pth}
\end{figure}

\section{Evolution of initial localized energy in a thermalized background}
\label{sec:evolution}
After thermalization, see Sec.\,\ref{sec:thermal_eq}, we can add to the system's variables the coordinates and momenta of a known breather, but it is simpler and more physical to add some kinetic energy to a single site so that the local energy becomes the intended breather energy $E_\text{b}$, and let the system evolve afterward. Then, we can consider that the system attains the thermal equilibrium when $P>N/2$ for the first time at $t_\text{th}$, which data we collect for later analysis in Sec.\,\ref{sec:qh}. The time value $t_\text{th}$ is also indicative of the maximal discrete breather lifetime in a given thermalized system. The participation number will continue changing with considerable fluctuation dispersion around some mean value close to $N/2$. Figure\, \ref{fig_brthermal1}-top shows the energy density contours and Fig.\,\ref{fig_brthermal1}-bottom shows the evolution of $P$ for $\Eloc=0.02$ in numerical simulations with two different breather energies $E_\text{b}=10 \Eloc$ and $E_\text{b}=21 \Eloc$, respectively, i.e., the energies of a single site excitations. Note that, in the second case, thermal equilibrium is not achieved during the observation time.

\subsection{Breather creation and statistics of the thermalization time}

To prevent the existence of two sources of large localization, we first localize the site with the largest local energy $e_n$ at site $n$ and if $E_\text{b}>e_n$, we change the momentum $p_n$ to $p_b$ such that the energy becomes the desired breather energy $E_\text{b}$, that is, $p_b=\sqrt{2(E_\text{b}-e_n)+p_n^2}$, with the sign of the original $p_n$ value. The coordinates are not changed.
After that, we leave the system to evolve until the first time $t_\text{th}$ for which $P>N/2$. The collected time $t_\text{th}$ depends on each particular realization of the numerical experiment. The standard deviation $\sigma$ of $t_\text{th}$ is very large, of the order of magnitude of its mean value, as it should be expected because breathers are complex structures that will not always be created with just the delivery of some kinetic energy to a single site. Some of the initial localized energy may be closer or further away from a breather or to breathers with different stability properties, which is also highly influenced by the initial background noise.
Despite that, the standard deviation of the mean $\sigma_m=\sigma/\sqrt{N_r}$, where $N_r$ is the number of measurements, becomes very small, indicating that the mean value of $t_\text{th}$ is a well-defined quantity. In all numerical simulations, we use $N_r=10^4$, but then discard the experiments where $t_\text{th}=0$, because the system is already found in the thermalized state at the beginning of a simulation, even after the addition of a single site excitation.

For some systems and low $E_\text{b}$, it is difficult to obtain nonzero $t_\text{th}$, and, therefore, a significant number of simulations are performed to obtain statistics where the average value of $t_\text{th}$ seems reasonable and $\sigma_m$ is small.  This problem is dealt with by multiplying the number of simulations many times. A more difficult problem for higher $E_\text{b}$ can be the extremely long life of some breathers, sometimes with the need for some days and many processors. Note that parallelization is only a partial solution to this problem as we have to wait until every random creation of a breather thermalizes, including the very stable ones, to prevent favoring realizations with shorter $t_\text{th}$ in the statistics.\footnote{The OMP directive NO WAIT at the end of the loop running the simulations cannot be used.}

\subsection{Exponential dependence of the thermalization time with respect to breather energies}
The dependence of the average thermalization time $\langle t_\text{th}\rangle$ as a function of the relative breather energies $E_\text{b}/\Eloc$ for the different systems are presented in the section below. Let us note here that, most likely, the small deviation from the exponential dependence observed at some curves at large energies can be attributed to the fact that the simulation time is eventually limited, and some very long simulations are excluded from the statistics.

Note that the exponential behavior of $t_{th}$ with respect to $E_\text{b}/\Eloc$ appears for sufficiently large values of $E_\text{b}$. This approximate threshold changes with the system and parameters, but it is about ten times the $\Eloc$ value. Indeed, in most cases, there is an initial diminution of the thermalization time. We conjecture that when the energy is not enough to make probable the formation of
breathers, the increase in momentum of a particle brings about the creation of phonons and the dissipation of localization. Only when the value of $E_\text{b}$ is large enough such that many
breathers are created
the exponential behavior consolidates.


\subsection{Dependence of the thermalization time with the participation number}
There is a huge variability of the thermalization time $t_\text{th}$ even for the same initial value of the participation number $P$. The particular values of $u_n$ and $p_n$ are of paramount importance for $t_\text{th}$. However, globally, as should be expected, for a given breather energy $E_\text{b}$, the larger the initial value of $P$, the shorter the thermalization time $t_\text{th}$, where the smallest linear correlation between $P$ and $t_\text{th}$ is found for the JJN system. Examples of the four systems are shown in Fig.\,\ref{fig_Pth}.

\section{Breather energies and thermalization times}
\label{sec:lifetime}
In this section, we present the four systems under study and analyze the dependence of the breather lifetimes on the breather energies, for different values of the coupling parameter $\kappa$. The preferred parameter for identifying breather energies is $E_\text{b}/\Eloc$, that is, the breather energy relative to the mean thermal energy. This is a logical parameter but the lifetimes are strongly dependent on the specific system and value of $\kappa$.

\subsection{System with quartic hard on-site potential and harmonic coupling (QH)}
\label{sec:qh}
The system with the hard quartic potential is described by Eq.\,\eqref{eq:quartic}, with $\omega_0=1$, the frequency of isolated oscillators, and the nonlinearity parameter $s=1$. The positive coupling constant $\kappa$ may take different values. Exact breathers, their obtention method, and their properties are presented in \ref{sec:app:qh}, particularly, their energies are of the order of a few tenths. We consider two values of the coupling parameter $\kappa=0.05$ and $\kappa=0.10$ and two values of the average initial local mean thermal energy $\Eloc=0.02$ and $\Eloc=0.04$. Breather energies are taken from $E_\text{b}=5\Eloc$ to $E_\text{b}=21\Eloc$. Smaller breather energies do not provide enough localization for breathers to form and larger values create exceptionally stable and long-lived breathers. Even during days of simulations on a supercomputer using one hundred processors, we do not achieve enough thermalization cases to provide good statistics. Generally speaking, we reproduce the method presented in the previous section $N_r=10^4$ times until thermalization is achieved. There is no time limit set for a simulation, but the supercomputer has a time limit of several days. This means that very long thermalization times (or infinite) are excluded from the statistics.

\begin{figure}[t]                                                                                                                                    \begin{center}
\includegraphics[width=0.6\textwidth]{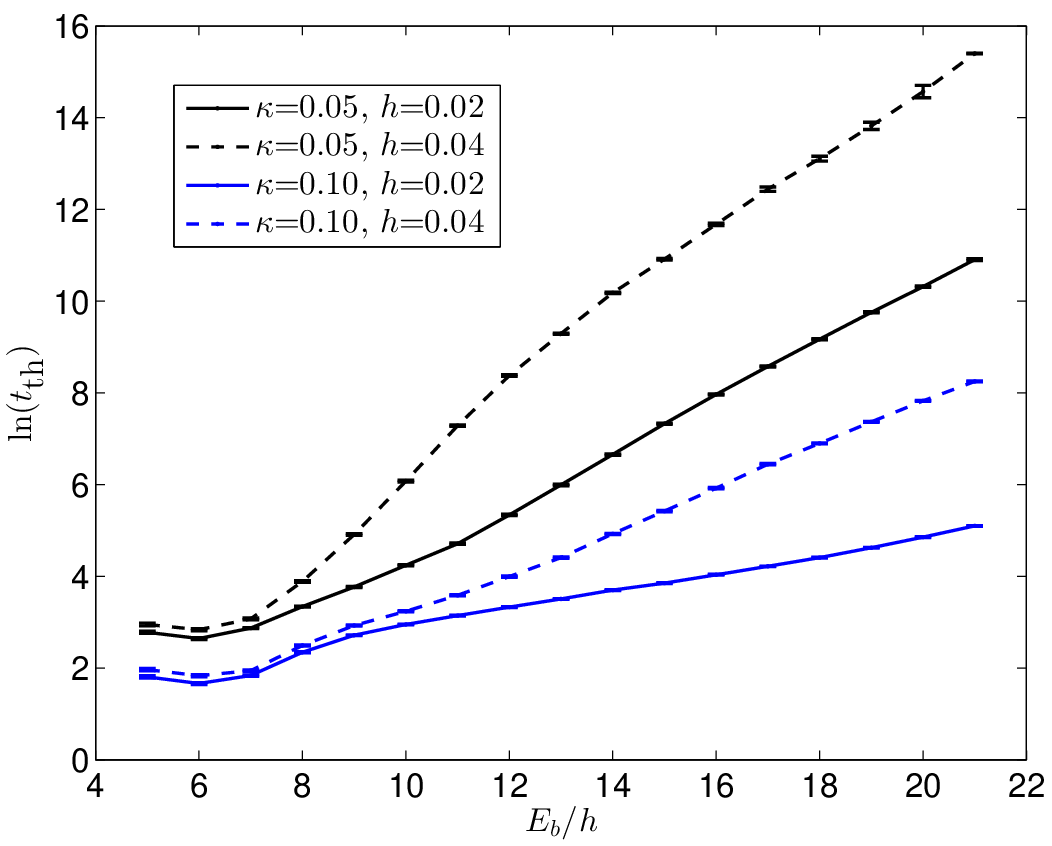}\,\,
\end{center}
\caption{Quartic hard system (QH): Thermalization time $t_\text{th}$ on a logarithmic scale as a function of the relative breather energy $E_\text{b}/\Eloc$ for coupling parameter values $\kappa=0.05$ (two upper curves) and $\kappa=0.10$ (two lower curves) and mean thermal energies $\Eloc=0.02$ (continuous lines) and $\Eloc=0.04$ (dashed lines). In this system, the breathers are exceptionally stable and the approximate exponential dependence for mid-to-large breather energies can be observed.
 }
\label{fig_qhtodas}
\end{figure}

Statistical results for the quartic hard potential case (QH) can be seen in Fig.\,\ref{fig_qhtodas}. The error bars are obtained by adding and subtracting the standard deviation of the mean, but they are very small and difficult to observe. In total, the results were obtained for seventeen different breather energy values. The exponential dependence of the thermalization time and averaged discrete breather lifetime can be easily seen and estimated, especially for larger breather energies, since the thermalization time $t_\text{th}$ axis is shown on a logarithmic scale. The obtained results can also be justified by the fact that larger energy breathers are more resilient toward background noise or thermal fluctuations. In addition, notice that the breather lifetime decreases as the coupling constant $\kappa$ increases, whereas the breather lifetime increases with the increase of the mean thermal energy value $\Eloc$.

\subsection{System with quartic soft on-site potential and harmonic coupling (QS)}
\label{sec:qs}
\begin{figure}[t]                                                                                                                                    \begin{center}
\includegraphics[width=0.60\textwidth]{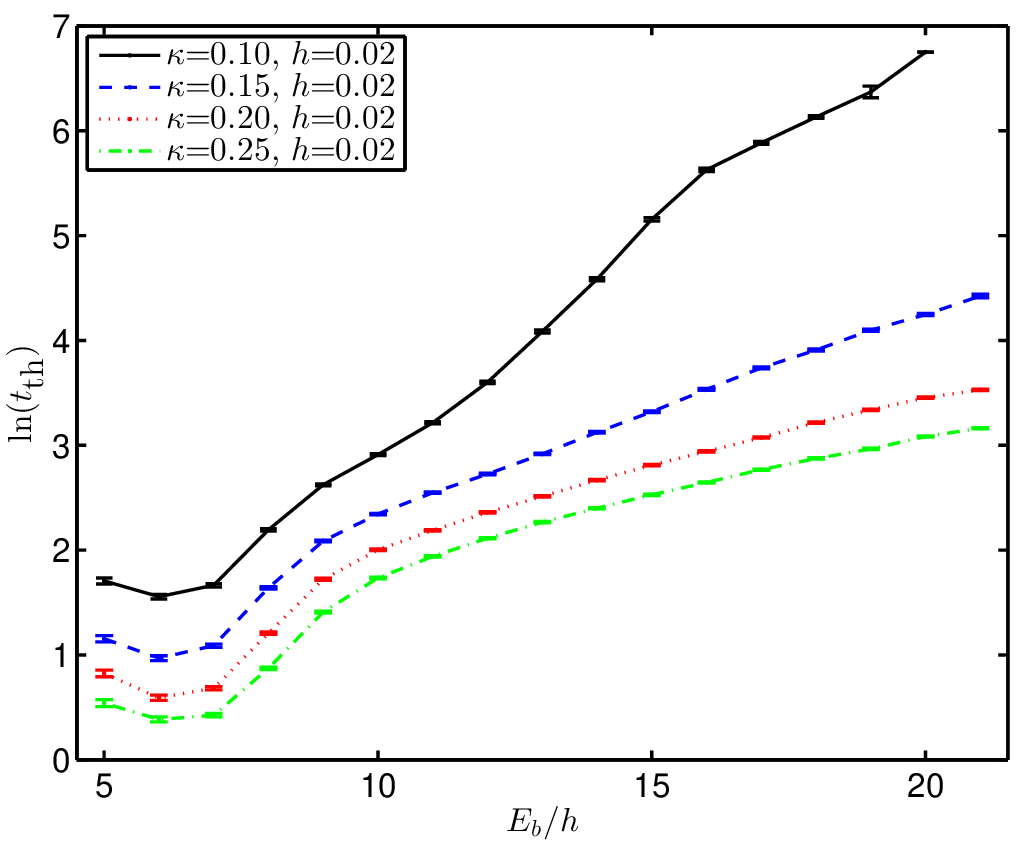}
\end{center}
\caption{  Quartic soft system (QS): Thermalization time $t_\text{th}$ on a logarithmic scale as a function of the relative breather energy $E_\text{b}/\Eloc$ for four different coupling parameter values $\kappa=0.10, 0.15, 0.20,0.25$, corresponding to the curves from top to bottom, and mean thermal energy $\Eloc=0.02$. The approximate exponential dependence for large breather energies is evident. See the text for an explanation. Note that the hump at the upper curve is a feature checked quite a few times with 20000 simulations per point. The reason is not known.
}
\label{fig_qstodas}
\end{figure}
The Hamiltonian in Eq.\,\eqref{eq:quartic} with $s=-1$ becomes soft. The on-site potential has a potential barrier at $u_n=\pm 1$ that separates the potential well centered at $u_n=0$ from a non-physical negative infinite potential well. The quartic soft on-site potential can be considered a reasonable approximation for a given system only inside the central potential well. Then, we write the code in such a way that if during a given simulation any $u_n$ leaves the safe well interval, the simulation is discarded. In this way, we are selecting a more distributed localization energy and diminishing the thermalization time. Therefore, the thermalization time cannot have an approximate exponential dependence on $E_\text{b}/h$.  For lower coupling $\kappa$, existing breathers localized at a single site have small energies, and as they are not created by an increase in $E_\text{b}$, the thermalization time $t_\text{th}$ increases linearly with $E_\text{b}/h$.  For a larger $\kappa$, when breathers have more energy and can be created, we observe again the approximate exponential dependence of $t_\text{th}$ as shown in Fig.\,\ref{fig_qstodas}. Results in Figure \ref{fig_qstodas} are shown only for a single mean thermal energy value while varying the coupling parameter $\kappa$ since for larger mean thermal energy, e.g., $\Eloc=0.04$, the increase in energy favors delocalization and particles are leaving the potential well towards a negative infinity well, with nonphysical results. Nevertheless, we are still observing a clear pattern that the average breather lifetime increases with the decreasing value of the coupling parameter $\kappa$ as already observed in the quartic hard potential case in Fig.\,\ref{fig_qhtodas}.

\subsection{Frenkel-Kontorova system with Lennard-Jones interaction potential}
\label{sec:fklj}
In this section, we consider a Frenkel-Kontorova (FK) system\,\cite{braun1998}, that is, with cosine on-site potential, and with the Lennard-Jones (LJ) interatomic interaction potential. This system provides a useful model for atoms in a crystal, taking into account the periodicity of the crystal. The on-site potential represents the interaction with other parts of the crystal, and the LJ potential provides a strong repulsion when two atoms approach each other and a potential well with a force that tends to zero when the atoms move apart, corresponding with the physical characteristics of interatomic interactions\,\cite{braun2004}. From a more technical point of view, the cosine potential provides a soft potential, without the nonphysical characteristics of the quartic soft potential of having an infinite negative well, and the need to continuously control that the coordinates do not penetrate into that well. It has also been used as a model for lattice excitations in silicates in 2D hexagonal lattices by Baj\=ars, Eilbeck, and Leimkhuler (BEL)\,\cite{bajars-physicad2015,bajars-quodons2015article,bajars2021}, and it has been shown that it is extremely easy to generate both stationary and moving breathers in one or two dimensions. Also, polarobreathers, i.e., breathers coupled to a charge, propagate in this system extremely well\,\cite{archilla-bajars2023}.

To compare with the quartic soft potential, we need an appropriate scaling as commented in  \ref{sec:app:fklj}. The Hamiltonian is given by:
\begin{align}
\begin{split}
H=\sum_n \Biggl(\frac{1}{2}p_n^2 +  &  U_0\left(1-\cos\left(2\pi\frac{u_n}{\sigma}\right)\right) \\
& +V_0\left[1+\frac{1}{\left(1+\fracc{u_{n+1}-u_n}{\sigma}\right)^{12}}-\frac{2}{\left(1+\fracc{u_{n+1}-u_n}{\sigma}\right)^6}\right] \Biggl)\,.
\label{eq:hamiltonianfklj}
\end{split}
\end{align}
As shown in \ref{sec:app:fklj}, we compare the FKLJ Hamiltonian with the QS, so as the linearized dynamical equations become identical. The result being that $U_0=\omega_0^2\sigma^2/(2\pi)^2$  and $\kappa=(\omega_0^2\sigma^2/72)V_0$. For the QS system, $\sigma=2$ and $\omega_0=1$, therefore, $U_0=1/\pi^2\simeq 0.103$ and $V_0=\kappa/18$.

\begin{figure}[t]                                                                                 \begin{center}
\includegraphics[width=0.7\textwidth]{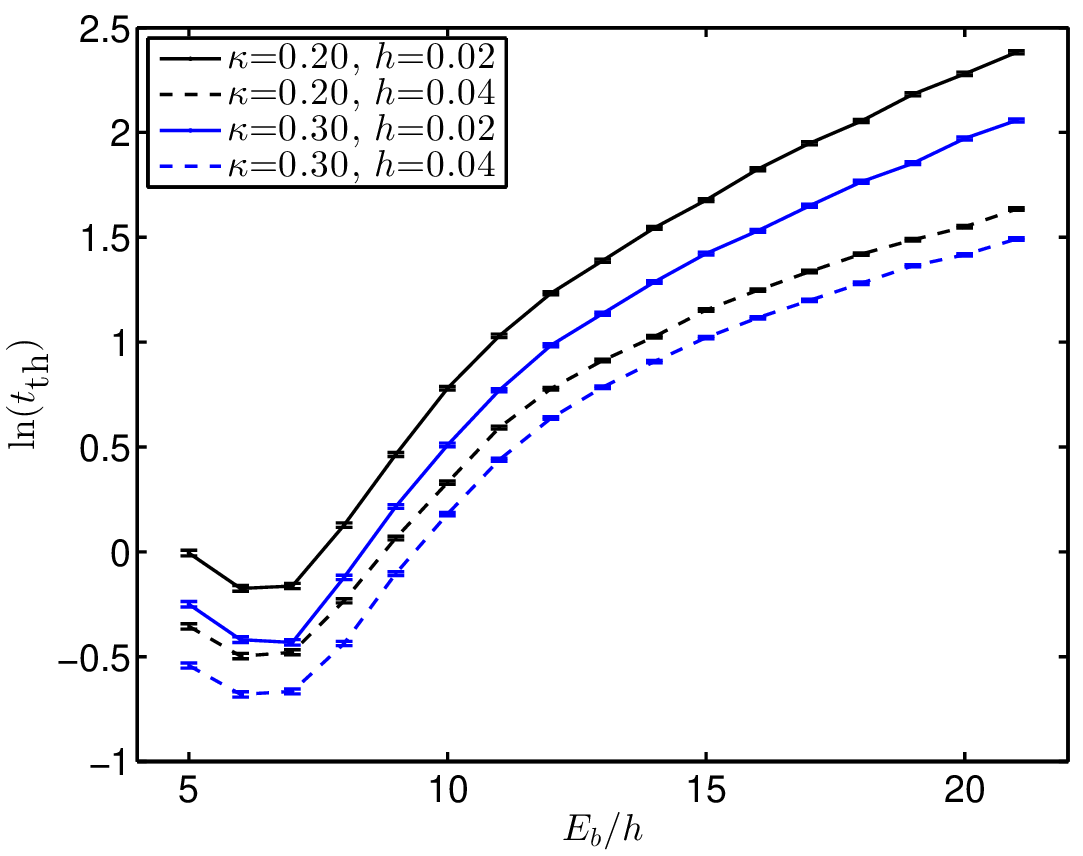}
\end{center}
\caption{ Frenkel-Kontorova and Lennard-Jones system (FKLJ): Thermalization time $t_\text{th}$ on a logarithmic scale as a function of the relative breather energy $E_\text{b}/\Eloc$ for coupling parameter values $\kappa=0.20$ (1st and 3rd from top) and $\kappa=0.30$ (2nd and 4th from top) and mean thermal energies $\Eloc=0.02$ (continuous line) and $\Eloc=0.04$ (dashed line). For smaller values of the coupling parameter $\kappa$, breathers have very small energy and they are destroyed when an extra energy $E_\text{b}$ is delivered. For relatively larger $\kappa$, breathers are formed and the exponential dependence of the thermalization time can be observed.
 }
\label{fig_fklj_todas}
\end{figure}

As this potential is soft and at low amplitudes is fitted with the quartic soft potential, there are similar features to be expected. In particular, for low coupling $\kappa=0.05$, the thermalization time $t_\text{th}$ shows a linear dependence with the delivered breather energy $E_\text{b}$, indicating that breathers are not formed because their energy is too low. The energy is rapidly dissipated. For $\kappa=0.20$ and $\kappa=0.30$, we again obtain the exponential dependence between $t_\text{th}$ and $E_\text{b}$, indicating that breathers are formed and long-lived in the thermalized system. Differently from the QS case, we can increase the temperature or the local mean thermal energy $\Eloc$, and $E_\text{b}$, without the control of $u_n$ going outside the infinite potential well. These results are shown in Fig.\,\ref{fig_fklj_todas}. Interestingly, thermalization times decrease as the coupling constant $\kappa$ increases but decrease as well when the mean thermal energy $\Eloc$ is increased, which is opposite to the quartic hard potential case and results in Fig.\,\ref{fig_qhtodas}.

There are other techniques to create breathers in both the QS and FKLJ systems consisting of adding and subtracting some momentum to a couple of neighboring particles, favoring the creations of $\pi$-breathers with energies above the phonon spectrum, but, in this paper, we limit our research only to a single-site excited breathers.

\subsection{Josephson Junction Network (JJN)}
\label{sec:jjn}

This system describes an array of Josephson Junctions (JJ), which is described by the Hamiltonian:
\begin{equation}
H=\sum_n \left(\fracc{1}{2}p_n^2+0.5\kappa\left(2-\cos(q_{n+1}-q_n)-\cos(q_n-q_{n-1})\right)\right)\,,   
\label{eq:Hjjn}
\end{equation}
Corresponding to a dynamical equation $\dot p_n=\ddot q_n=-\partial H/\partial q_n$:
\begin{equation}
\ddot q_n= \kappa(\sin(q_{n+1}-q_n)-\sin(q_{n}-q_{n-1})). 
\label{eq:dynamicjjn}
\end{equation}
The variable $q_n$ is written differently, as it is an angle variable representing the phase of the superconducting order parameter\,\cite{barone1982,lando-flach2023}. The {\em kinetic} energy $\fracc{1}{2}p_n^2$ is the island Coulomb charging energy, and $\kappa$ is the Josephson coupling between two neighboring superconducting islands. It also represents  the angle of a rotating coupled pendulum, which allows for an easier intuition of the phenomena.
This system has no on-site potential, which makes it a very different system from the previous three. For example, the phonon band is not bounded from below and extends from $\omega=0$ to $\omega=2\sqrt{\kappa}$ as shown in \ref{sec:app:jjn}.   When increasing a momentum $p_n$ for breather creation it is convenient to subtract an appropriate amount to all rotators to prevent a global rotation of the system, which is equivalent to set to zero the mean electric potential of the JJ network.

\subsubsection{Breathers in JJN}
\label{sec:breathersJJN}
In this system, exact single-site breathers do not exist as shown in \ref{sec:app:jjn}, but there are long-lived transient localized entities, which are formed during some time. The evolution to thermal equilibrium can be seen in Fig.\,\ref{fig_etn_rotk05h02}-left. The localization corresponds physically to a strong AC component of the superconducting currents across two neighboring junctions, as shown in Fig.\,\ref{fig_etn_rotk05h02}-right and explained below.

\subsubsection{Particularities of thermalization in JJNs}
\label{sec:thermalJJN}
Interestingly, the numerical thermalization procedure leads to a value of $\langle P\rangle$  17\% above $N/2$, that is, about $5-6$ particles for $N=64$. This percentage continues when increasing the system size to 128 or 256. The reason is that the potential energy is localized completely at the bonds, and then shared always between two particles. If we assign the bond energy to a single particle, that is, $E_n=\fracc{1}{2}p_n^2+\kappa(1-\cos(q_{n+1}-q_n))$, then the thermalized $P$ becomes almost exactly $N/2$. Note that in the approximate deduction of $\langle P \rangle$, we assumed a description of the particles with energy $e_n$ relatively independent of the neighbors. This hypothesis holds quite well for the systems with on-site potential but it is clearly not valid for FPUT systems where the potential energy is at the bonds.  A different test is adding an on-site potential $U(u_n)=\omega_0^2(1-\cos(q_n))$ with $\omega_0=1$. In this case, the value of $\langle P\rangle $ at thermal equilibrium is again slightly above $N/2$.
However, even for the FPUT system the value of $P=N/2$ is well within the oscillations of $P$ at thermal equilibrium, indicating that the system is fast approaching it. This can be confirmed by changing the thermalization condition to $P>N/2+6$ or about the observed $\langle P \rangle$. The differences are only apparent at low breather energies $E_\text{b}$ before the exponential behavior takes place.

\begin{figure}[t]                                                                    \begin{center}\includegraphics[width=0.49\textwidth]{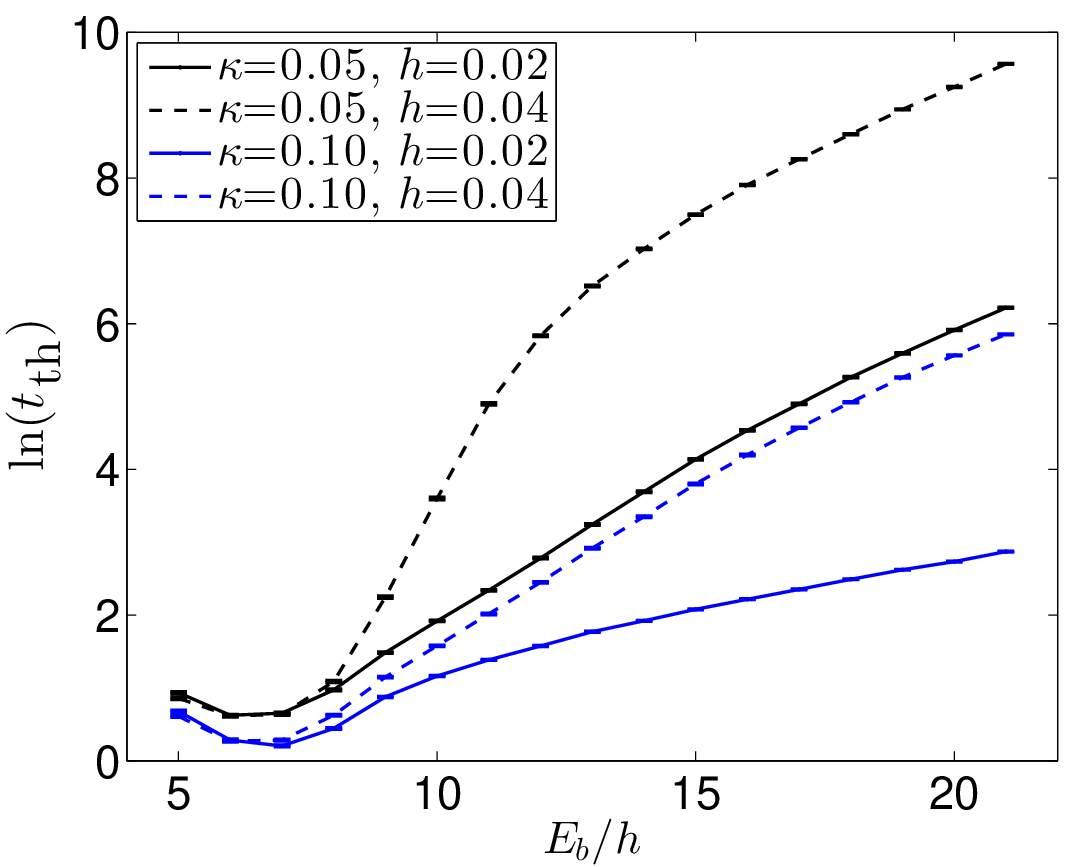}\,
\includegraphics[width=0.49\textwidth]{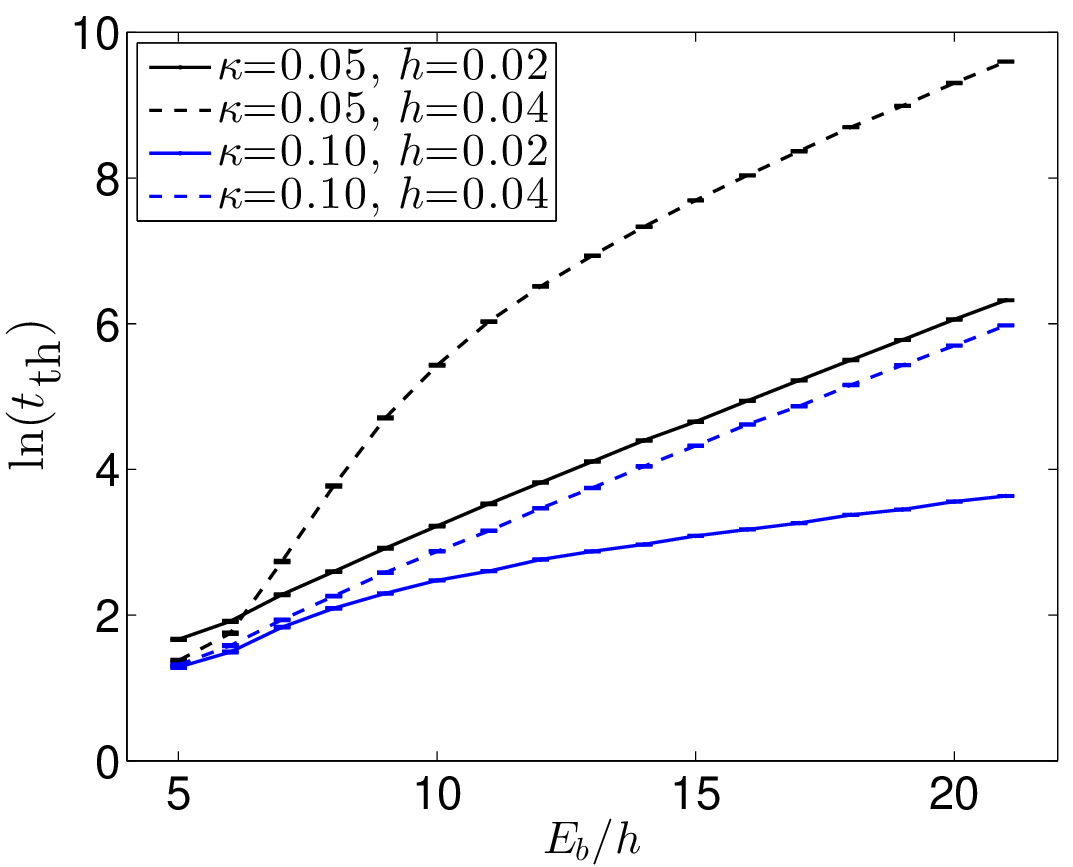}
\end{center}
\caption{ ({\bf Left}) Josephson junction network (JJN): Using thermalization condition $P>N/2$, thermalization time $t_\text{th}$ on a logarithmic scale as a function of the relative excitation energy $E_\text{b}/\Eloc$ for coupling parameter values $\kappa=0.05$ and (two upper curves) and $\kappa=0.10$ (two lower curves) and mean thermal energies $\Eloc=0.02$ (continuous lines) and $\Eloc=0.04$ (dashed lines). The exponential behavior for sufficiently large energies is testimony to the nonlinear localization of energy.
({\bf Right}) Same plot but results were obtained for thermalization condition $P>N/2+6$. Differences are only at small $E_\text{b}$ as expected. See text.
 }
\label{fig_rot_todas}
\end{figure}

Despite the differences and for similar values of the parameters, this system also shows exponential dependence of the thermalization time for the localized energy delivered, as shown in Fig.~\ref{fig_rot_todas}. Compared to other systems, Fig.\,\ref{fig_qhtodas}--\ref{fig_fklj_todas}, for the JJN system the thermalization time $t_\text{th}$ increases with increasing coupling constant $\kappa$ while decreases with decreasing mean thermal energy $\Eloc$.

\subsubsection{The physical meaning of localization in JJNs}
\label{sec:meaningJJN}

Let us comment on the physical meaning of the localization for JJNs: the two fundamental equations of the junction are\,\cite{barone1982}:
\begin{align}
I_{n+1,n}&=I_c\sin(\phi_{n+1,n})  \quad &\mathrm{and} \quad {\cal V}_{n+1,n}&=\fracc{\hbar}{2e}\Omega_{n+1,n}, \quad \mathrm{with}\\
\phi_{n+1,n}&=q_{n+1}-q_n    \quad &\mathrm{and} \quad \Omega_{n+1,n}&=\fracc{\d}{\d t}\phi_{n+1,n}=p_{n+1}-p_n\, ,
\end{align}
\noindent
where $I_{n+1,n}$ is the superconducting current across the junction between two superconducting islands (SCI) $n$ and $n+1$, $\phi_{n+1,n}$ is the different in phase between the two SCI, ${\cal V}_{n+1,n}$ is the potential difference between SCI, and $\Omega_{n+1,n}$ is the instantaneous frequency of the phase difference between junctions. The critical current $I_c$ depends on the particular junction, while $\hbar$ and $e$ are physical constants, but in what follows, we will just use $I_c=1$ and $\hbar/2e=1$.

If the potential difference across a junction is zero, then $\Omega_{n+1.n}=0$, $\phi_{n+1,n}$ is constant, and $I_{n+1,n}$ is a DC current. This is called the DC Josephson junction effect (DC JJE).

However, if the potential and the frequency $\Omega_{n+1,n}$ are constant, the SC current becomes an AC current with that frequency. This is called the AC Josephson junction effect (AC JJE)

\begin{figure}[t]
\begin{center}
\includegraphics[width=0.53\textwidth]{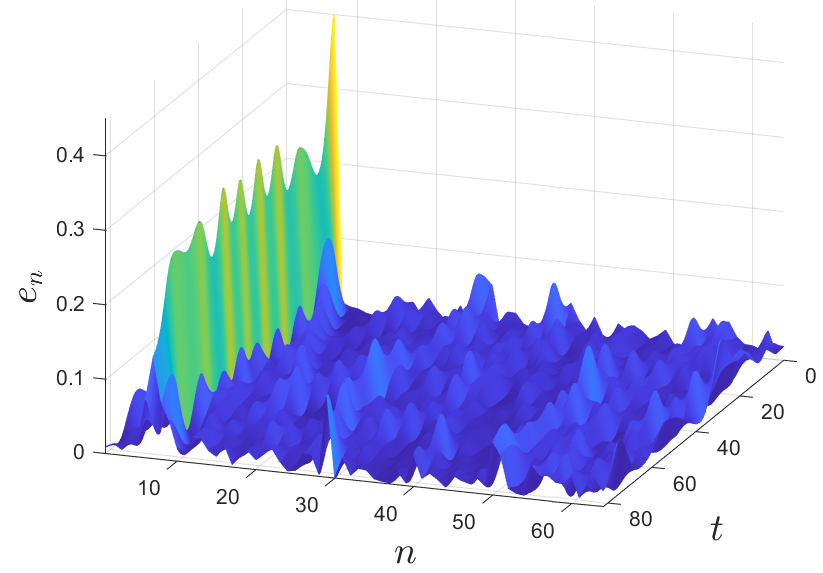}
\includegraphics[width=0.46\textwidth]{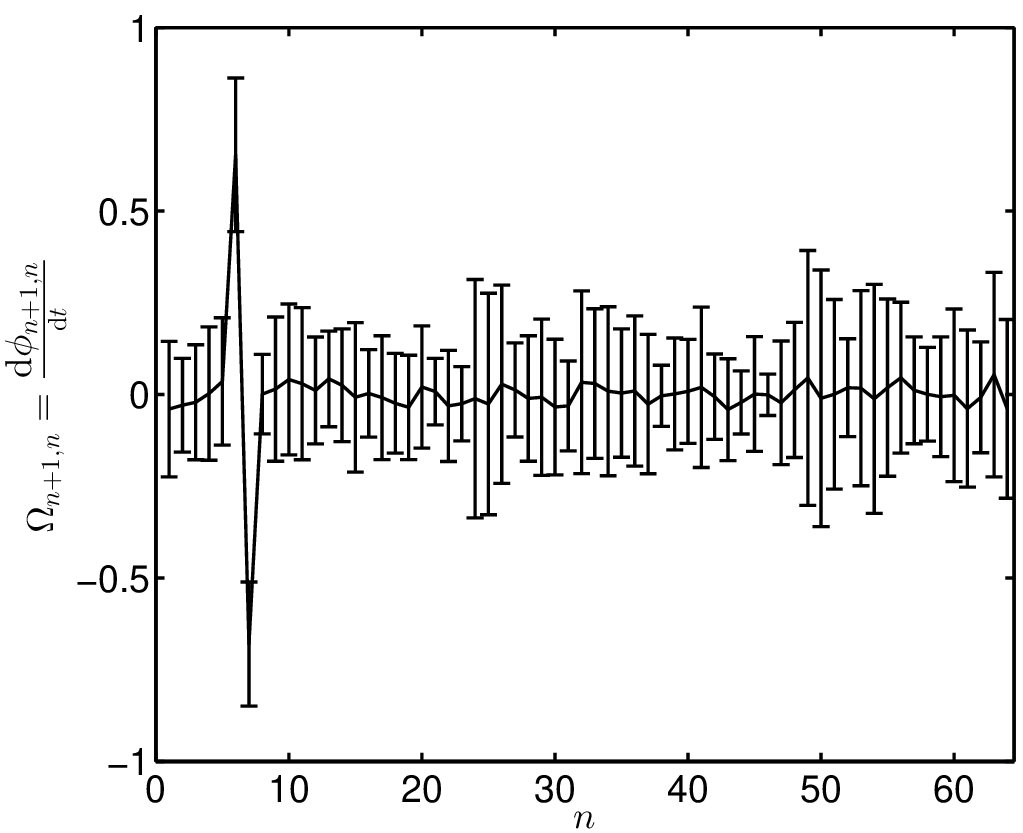}\\
\includegraphics[width=0.46\textwidth]{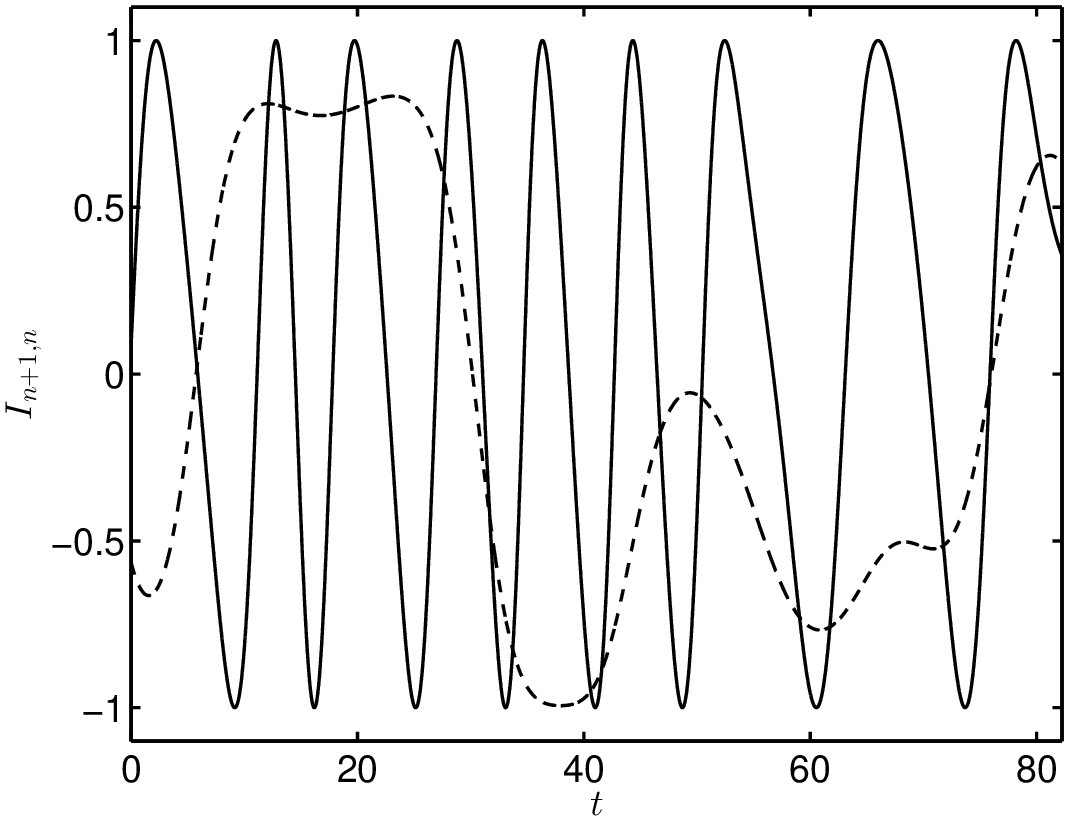}
\end{center}
\caption{({\bf Left}:) Josephson junction network (JJN): Example of the route to thermal equilibrium for $\kappa=0.05$, $\Eloc=0.02$, for a delivered energy of $E_\text{b}=0.40$ with $P=13$. This is a short thermalization time example, where there are only a few localized oscillations before thermalization. ({\bf Right}:) Time average value with the standard deviation of  ${\Omega}_{n+1,n}=\frac{\d \phi_{n+1,n}}{\d t}$ across each junction. The averaging interval excludes the formation and decaying intervals. ({\bf Bottom}:) Time dependence of the supercurrent $I=I_c\sin(\phi_{n+1,n})$ across one the two excited junctions and another one three sites apart. Other simulations are similar but with thousands of oscillations.
}
\label{fig_etn_rotk05h02}
\end{figure}

In our system the localization appears as a SCI where $\dot q_n$ is large with a definite sign and oscillations smaller than its value. The frequency of the phase differences across the two neighboring junctions is then well defined and with opposite signs as can be seen in Fig.\,\ref{fig_etn_rotk05h02}-right and, therefore, these two junctions experience the JJ AC effect, having well-defined frequencies larger than the other junctions. The rest of the junctions experience badly defined and changing frequencies smaller than the AC junctions. This is illustrated in Fig.\,\ref{fig_etn_rotk05h02}-bottom for a short life breather. For longer-lived breathers, the effect is similar but with thousands of oscillations. This short-lived breather has been chosen so that Fig.\,\ref{fig_etn_rotk05h02}-left can easily be seen in its entirety.

To conclude, for the JJN system, the localization appears not as the amplitude of the SC current, which is bounded by $I_c$, but by localization in frequency. For the pendula chain analog, the localization is also an angular frequency with a definite sign, that is a rotating pendulum.

\section*{Conclusions}
We have explored the route to thermalization in different systems of oscillators, corresponding to a variety of physical systems. We have introduced a parameter, the participation number $P$, that measures the degree of localization of energy in a given system, taking values between 1 and $N$, the particle number. We have deduced through an approximate method that its value at thermal equilibrium is $N/2$, and observed in simulations that, although, not exact, this condition indicates that the system is very close to thermal equilibrium. We have developed a method to create an initial thermalized system and to deliver a defined amount of localized energy, observing the route to thermalization afterward.
For values of the parameters for which breathers exist, it is observed that the thermalization time has an approximate exponential dependence on the breather energy, if this  energy is larger than  approximately ten times the local average thermal energy. If breathers do not exist or are unstable, the route to equilibrium shows a linear dependence on the thermalization time.
In some cases, as the quartic hard potential, breathers are extremely stable, and the thermalization time poses problems to powerful computers, including a supercomputer cluster, which eschews the statistics for large breather energies to shorter thermalization times.

To conclude, the existence of breathers in a system has a measurable consequence, an approximate exponential relaxation time to equilibrium. Their long life may prevent the evacuation of heat in environments where they are created in huge numbers, such as fusion reactors.

\section*{Acknowledgements}
JFRA acknowledges the Center for Theoretical Physics of Complex Systems at the Institute of Basic Science in Daejeon, Republic of Korea, for hospitality.

\section*{Funding}
JFRA thanks grant PID2022-138321NB-C22 funded by MICIU/AEI/10.13039/501100011033 and ERDF/EU, and travel help both from Universidad de Sevilla VIIPPITUS-2024, and PCS at the Institute of Basic Science.

JB acknowledges financial support from the Faculty of Science and Technology of the University of Latvia.

SF acknowledges the financial support from the Institute for Basic Science (IBS) in the Republic of Korea through the Project No. IBS-R024-D1.

\section*{Computational resources}
The computational resources have been provided by a dedicated computer Intel Core i7-12700KF with 12 processors of 12th generation at 3.60 GHz with 64 GB RAM, managed by JB at the University of Latvia, and by the use of H\'ercules, the supercomputer of the Scientific Computing Center of Andaluc\'ia (CICA), where up to a thousand processors have been used for parallel processing.

\section*{References}

\begin{thebibliography}{10}
\expandafter\ifx\csname url\endcsname\relax
  \def\url#1{\texttt{#1}}\fi
\expandafter\ifx\csname urlprefix\endcsname\relax\def\urlprefix{URL }\fi
\expandafter\ifx\csname href\endcsname\relax
  \def\href#1#2{#2} \def\path#1{#1}\fi

\bibitem{mackayaubry94}
R.~S. MacKay, S.~Aubry, Proof of existence of breathers for time-reversible or
  {H}amiltonian networks of weakly coupled oscillators, Nonlinearity 7 (1994)
  1623.

\bibitem{flach1998}
S.~Flach, C.~R. Willis, Discrete breathers, Phys. Rep. 295 (1998) 181--164.

\bibitem{2004PhT....57a..43C}
D.~K. {Campbell}, S.~{Flach}, Y.~S. {Kivshar}, Localizing energy through
  nonlinearity and discreteness, Physics Today 57~(1) (2004) 43--50.

\bibitem{flach2008}
S.~Flach, A.~V. Gorbach, Discrete breathers. {A}dvances in theory and
  applications, Phys. Rep. 467~(1-3) (2008) 1--116.

\bibitem{sievers-takeno1988}
A.~J. Sievers, S.~Takeno, {Intrinsic localized modes in anharmonic crystals},
  Phys. Rev. Lett. 61 (1988) 970--973.

\bibitem{anderson1958}
P.~W. Anderson, Absence of diffusion in certain random lattices, Phys. Rev. 109
  (1958) 1492--1505.

\bibitem{archilla1999}
J.~F.~R. Archilla, R.~S. Mackay, J.~L. Mar\'in, Discrete breathers and
  {A}nderson modes: two faces of the same phenomenon?, Physica D 134 (1999)
  406--418.

\bibitem{flach1995}
S.~Flach, Obtaining breathers in nonlinear {H}amiltonian lattices, Phys. Rev. E
  51~(4) (1995) 3579--3587.

\bibitem{marinaubry96}
J.~L. Mar\'{\i}n, S.~Aubry, Breathers in nonlinear lattices: Numerical
  calculation from the anticontinuous limit, Nonlinearity 9 (1996) 1501.

\bibitem{marin1998}
J.~L. Mar\'{\i}n, J.~C. Eilbeck, F.~M. Russell, Localized moving breathers in a
  {2D} hexagonal lattice, Phys. Lett. A 248~(2-4) (1998) 225--229.

\bibitem{archilla2001}
J.~F.~R. Archilla, P.~L. Christiansen, S.~F. Mingaleev, Y.~B. Gaididei,
  Numerical study of breathers in a bent chain of oscillators with long-range
  interaction, J. Phys. A. Math. Gen. 34~(33) (2001) 6363--6373.

\bibitem{itertokamak2023}
{ITER}, Tokamak, {\tt https://www.iter.org/mach/tokamak}, accesed, Nov 8
  (2024).

\bibitem{russell-archilla2024}
F.~M. Russell, J.~F.~R. Archilla, J.~L. Mas, Quodon current in tungsten and
  consequences for tokamak fusion reactors, Phys. Status Solidi RLL 18~(2)
  (2024) 2300297.

\bibitem{IvanchenkoDiscretePhysicaD2004}
M.~V. Ivanchenko, O.~I. Kanakov, V.~D. Shalfeev, S.~Flach, Discrete breathers
  in transient processes and thermal equilibrium, Physica D 198~(1-2) (2004)
  120--135.

\bibitem{khadeeva-dmitriev2011}
L.~Z. Khadeeva, S.~V. Dmitriev, Lifetime of gap discrete breathers in diatomic
  crystals at thermal equilibrium, Phys. Rev. B 84 (2011) 144304.

\bibitem{riviere-piazza2019}
A.~Rivi\`ere, S.~Lepri, D.~Colognesi, F.~Piazza, Wavelet imaging of transient
  energy localization in nonlinear systems at thermal equilibrium: {T}he case
  study of {NaI} crystals at high temperature, Phys. Rev. B 99~(2) (2019)
  024307.

\bibitem{iubini2019}
S.~Iubini, L.~Chirondojan, G.-L. Oppo, A.~Politi, P.~Politi, Dynamical freezing
  of relaxation to equilibrium, Phys. Rev. Lett. 122 (2019) 084102.

\bibitem{rasmussen00}
K.~O. Rasmussen, T.~Cretegny, P.~G. Kevrekidis, N.~Gr\o{}nbech-Jensen,
  Statistical mechanics of a discrete nonlinear system, Phys. Rev. Lett. 84
  (2000) 3740--3743.

\bibitem{rumpf2008}
B.~Rumpf, Transition behavior of the discrete nonlinear {S}chr{\"{o}}dinger
  equation, Phys. Rev. E 77~(3) (2008) 036606.

\bibitem{rumpf09}
B.~Rumpf, Stable and metastable states and the formation and destruction of
  breathers in the discrete nonlinear {S}chr{\"o}dinger equation, Physica D
  238~(20) (2009) 2067 -- 2077.

\bibitem{PhysRevLett.120.184101}
T.~Mithun, Y.~Kati, C.~Danieli, S.~Flach, Weakly nonergodic dynamics in the
  {G}ross-{P}itaevskii lattice, Phys. Rev. Lett. 120 (2018) 184101.

\bibitem{ChernyNOn-GibbsPhysicalReviewA2019}
A.~{\relax Yu}. Cherny, T.~Engl, S.~Flach, Non-{G}ibbs states on a
  {B}ose-{H}ubbard lattice, Phys. Rev. A 99~(2) (2019) 023603.

\bibitem{DanieliCampbellFlachPRE2017}
C.~Danieli, D.~K. Campbell, S.~Flach, Intermittent many-body dynamics at
  equilibrium, Phys. Rev. E 95~(6) (2017) 060202.

\bibitem{braun1998}
O.~M. Braun, {\relax Yu}.~S. Kivshar, Nonlinear dynamics of the
  {F}renkel-{K}ontorova model, Phys. Rep. 306 (1998) 1--108.

\bibitem{braun2004}
O.~M. Braun, {\relax Yu}.~S. Kivshar, The {F}renkel-{K}ontorova Model:
  Concepts, Methods, and Applications, Springer, Berlin-Heidelberg, 2004.

\bibitem{lennard-jones1925}
J.~E. Lennard-Jones, On the forces between atoms and ions., P. R. Soc. A
  109~(752) (1925) 584--597.

\bibitem{lennard-jones1929}
J.~E. Lennard-Jones, The electronic structure of some diatomic molecules., T.
  Faraday Soc. 25 (1929) 0668--0685.

\bibitem{schwerdtfeger2024}
P.~Schwerdtfeger, D.~J. Wales, 100 years of the {L}ennard-{J}ones potential, J.
  Chem. Theory Comput. 20~(9) (2024) 3379--3405.

\bibitem{barone1982}
A.~Barone, G.~Paterno, Physics and Applications of the {J}osephson Effect, John
  Wiley \& Sons, New York, 1982.

\bibitem{lando-flach2023}
G.~M. Lando, S.~Flach, Thermalization slowing down in multidimensional
  {J}osephson junction networks, Phys. Rev. E 108 (2023) L062301.

\bibitem{reif2009}
F.~Reif, Fundamentals of statistical and thermal physics, Waveland Press, Long
  Grove, 2009.

\bibitem{schwabl2006}
F.~Schwabl, Statistical Mechanics, 2nd Edition, Springer, Berlin-Heidelberg,
  2006.

\bibitem{bajars-physicad2015}
J.~Bajars, J.~C. Eilbeck, B.~Leimkuhler, Nonlinear propagating localized modes
  in a {2D} hexagonal crystal lattice, Physica D 301-302 (2015) 8 -- 20.

\bibitem{bajars-quodons2015article}
J.~Bajars, J.~C. Eilbeck, B.~Leimkuhler, Numerical simulations of nonlinear
  modes in mica: Past, present and future, Springer Ser. Mater. Sci. 221 (2015)
  35--67.

\bibitem{bajars2021}
J.~Baj\=ars, J.~C. Eilbeck, B.~Leimkuhler, 2{D} mobile breather scattering in a
  hexagonal crystal lattice, Phys. Rev. E 103 (2021) 022212.

\bibitem{archilla-bajars2023}
J.~F.~R. Archilla, J.~Baj\={a}rs, Spectral properties of exact polarobreathers
  in semiclassical systems, Axioms 12 (2023) 437.

\bibitem{aubry2006}
S.~Aubry, Discrete breathers: Localization and transfer of energy in discrete
  {H}amiltonian nonlinear systems, Physica D 216 (2006) 1--30.

\bibitem{ford_FPU1992}
J.~Ford, The {F}ermi-{P}asta-{U}lam problem --{P}aradox turns discovery, Phys.
  Rep. 213~(5) (1992) 271--310.

\bibitem{sanchez-rey2024}
B.~S\'anchez-Rey, G.~James, J.~Cuevas, J.~F.~R. Archilla, Bright and dark
  breathers in {F}ermi-{P}asta-{U}lam lattices, Phys. Rev. B 70~(014301) (2024)
  1--10.

\end{thebibliography}
\newcommand{\noopsort}[1]{} \newcommand{\printfirst}[2]{#1}
  \newcommand{\singleletter}[1]{#1} \newcommand{\switchargs}[2]{#2#1}

\appendix

\section{Breathers from the anticontinuous limit}
\label{sec:app:anticountinuous}
\setcounter{figure}{0}
We first construct breathers with a given frequency $\omega_b$ at the anticontinuous limit, that is, with zero coupling $\kappa=0$. They consist of one or various isolated excited oscillators while the rest of the oscillators are at rest.
\subsection{Anticontinuous limit}
The details of the technique have been explained in detail in different publications\,\cite{flach1995,marinaubry96,aubry2006}. We construct numerically the Fourier components $z_k$ of the single oscillator of a given frequency, using as a seed $z_k=0.3$ of values of that order obtaining the initial coordinates. For stationary breathers, we can fix the initial phase by choosing  $p_n(0)=0$. Then the system is time-reversible and the solution becomes determined by their initial position. The solution becomes $u_n(t)=z_{0,n}+\sum_{k=1}^{k_m} 2 z_{k,n}\cos(k\omega_b t)$, with $k_m$ an arbitrary maximum value of the harmonic order, that we take as $k_m=15$.  By path continuation, we can obtain the initial coordinate as a function of the frequency, i.e., $u_0=u_0(\omega_b)$. We can also obtain the Floquet eigenvalues, which are trivial but useful. There are two eigenvalues at $+1$, corresponding to the phase mode and growth mode of the isolated oscillator with frequency $\omega_b$. They are due to the proximity of a solution with slightly different phase and frequency, respectively. The rest of the eigenvalues correspond to small amplitude perturbation of the oscillators at rest, which produce harmonic oscillations with frequency $\omega_0$ given by $u_n=\exp(\pm \ii \omega_0 t)$, with Floquet eigenvalue $\exp(\pm\ii \omega_0 T_b)=\exp(\pm\ii 2\pi\omega_0/\omega_b)$. They will be at angles $\pm 327^\circ=\mp 33^\circ $ for $\omega_b=1.1$ and  $\pm 240^\circ=\mp 120^\circ$ for $\omega_b=1.5$ as can be seen in Fig.\,\ref{fig_qhsinglewb}-right.

When we connect the oscillators with $\kappa>0$, the linear modes will have frequencies $\omega=\sqrt{\omega_0^2+4\kappa\sin^2(q/2)}$, with $q$, the wavenumber or momentum. That is, they will be between frequencies $\omega_0$ and $\omega_{max}=\sqrt{\omega_0^2+4\kappa}$, with Floquet eigenvalues $\exp(\ii 2\pi \omega/\omega_b)$ approaching towards $+1$ where they may produce an instability as they will have the same frequency as the breather and therefore they will be excited.

Using the initial isolated solution, we can have just two possibilities for every single time-reversible oscillator coded with the signature $\sigma=1$ if $u_0>0$ and $\sigma=-1$ if $u_0<0$. Single breathers are obtained for the signature $[0...0 1 0 0...0]$.

\begin{figure}[htb]
\begin{center}
\includegraphics[width=0.49\textwidth,height=0.49\textwidth]{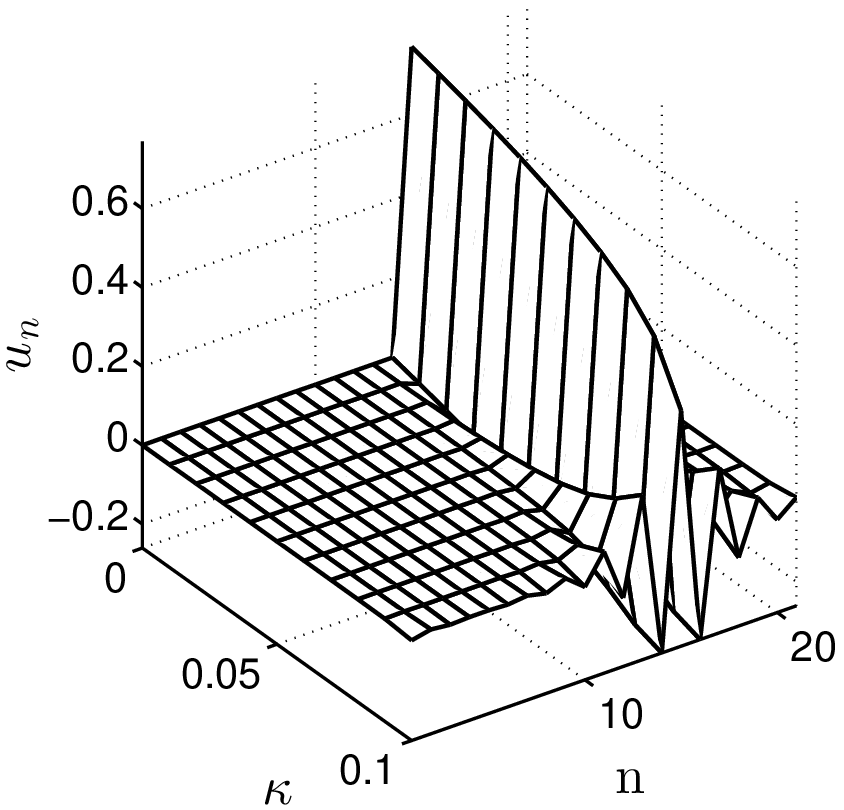}\,\,
\includegraphics[width=0.49\textwidth,height=0.49\textwidth]{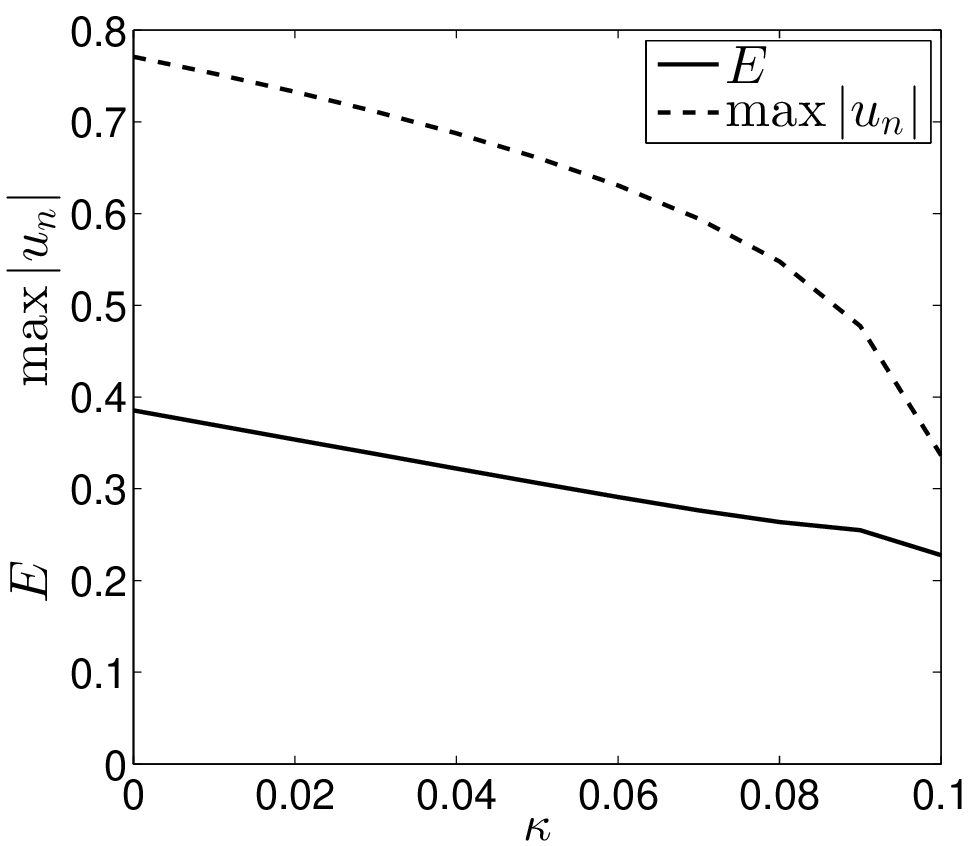}
\end{center}
\caption{System with quartic hard potential (QH):  ({\bf Left:}) Profile of breathers with $\omega_b=1.2$ as a function of the coupling parameter $\kappa$. For $\kappa=1.1$, the breather becomes unstable. ({\bf Right:}) Maximum value of the initial condition (dashed line) and energy (continuous line) as functions of the coupling parameter $\kappa$.
 }
\label{fig_brh12s1profile}
\end{figure}

\section{System with hard quartic potential and harmonic coupling (QH)}
\label{sec:app:qh}
\setcounter{figure}{0}
We consider both symmetric hard and soft Klein-Gordon potentials and harmonic coupling. Then, keeping only the first nonlinear term of the on-site potential, the Hamiltonian of the system is given by:
\begin{equation}
H=\sum_n \fracc{p_n^2}{2 m}+ m\omega_0^2\left(\frac{u_n^2}{2}+s \frac{u_n^4}{4}\right) +\kappa \fracc{1}{2}(u_{n+1}-u_n)^2\, ,
\label{eq:Hphys}
\end{equation}
where $\omega_0$ is the frequency of the isolated oscillator at the linear limit. Initially, we consider the hard potential, so $s=+1$.

We can re-scale the system with units $u_L=a$ the lattice distance, $u_T=1/\omega_0$, and $u_M=m$, then, the scaled equation is given by the Hamiltonian \eqref{eq:quartic}:
\begin{equation}
H=\sum_n \fracc{p_n^2}{2}+ \omega_0^2\left(\frac{u_n^2}{2}+s \frac{u_n^4}{4}\right) +\kappa \fracc{1}{2}(u_{n+1}-u_n)^2\, ,
\label{eq:Hscaled2}
\end{equation}
where we use the same symbols for the scaled variables. The scaled isolated linear frequency is $\omega_0=1$, although we keep the symbol to keep its meaning explicit and such that it is easy to compare with other scaling. The value of $u_n$ should be generally speaking sufficiently smaller than unity, which is now the lattice distance. Following the procedure explained below, we find that for hard breathers,  the nonlinearity parameter $s=1$ produces amplitudes of about 0.6 for a frequency $\omega_b=1.2$, which seems reasonable. Note that a change in the distance scale in $\bar u=a u$ is equivalent to a change in the parameter to $\bar s=a^2 s$, so $s$ can always be chosen to be the unity.

 \begin{figure}[t]
\begin{center}
\includegraphics[width=0.49\textwidth]{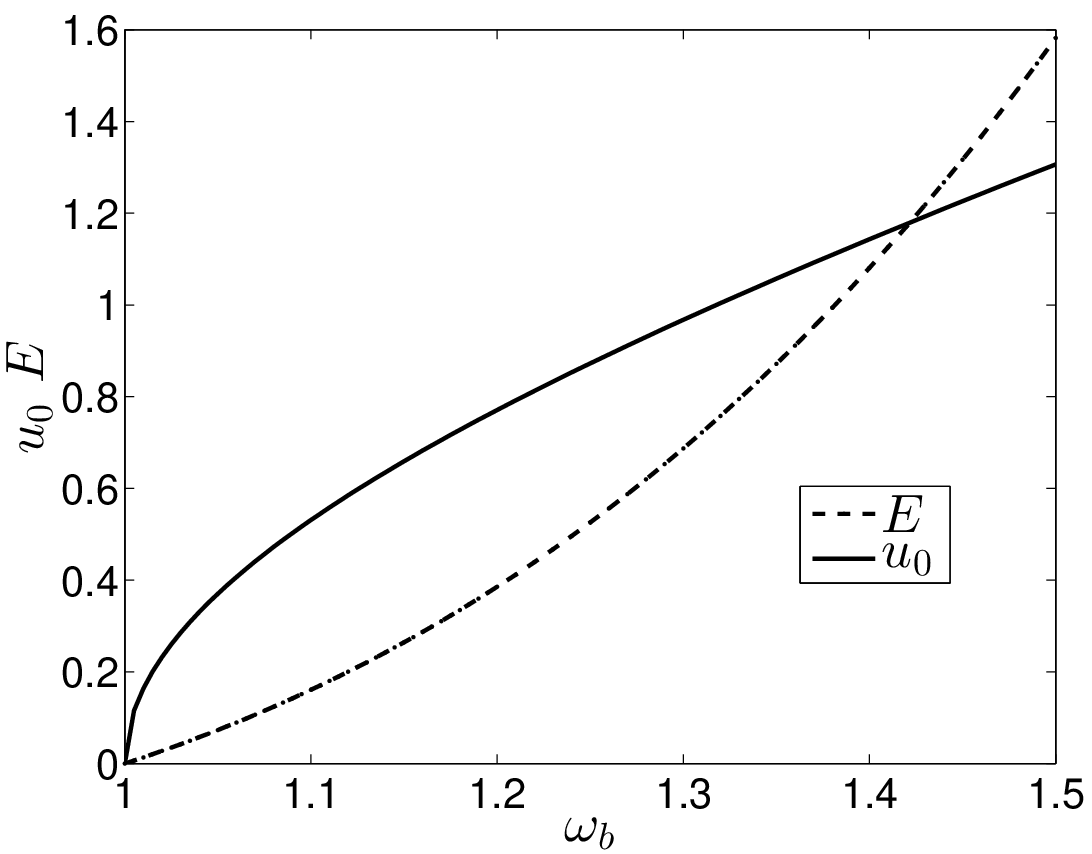}\,\,
\includegraphics[width=0.49\textwidth]{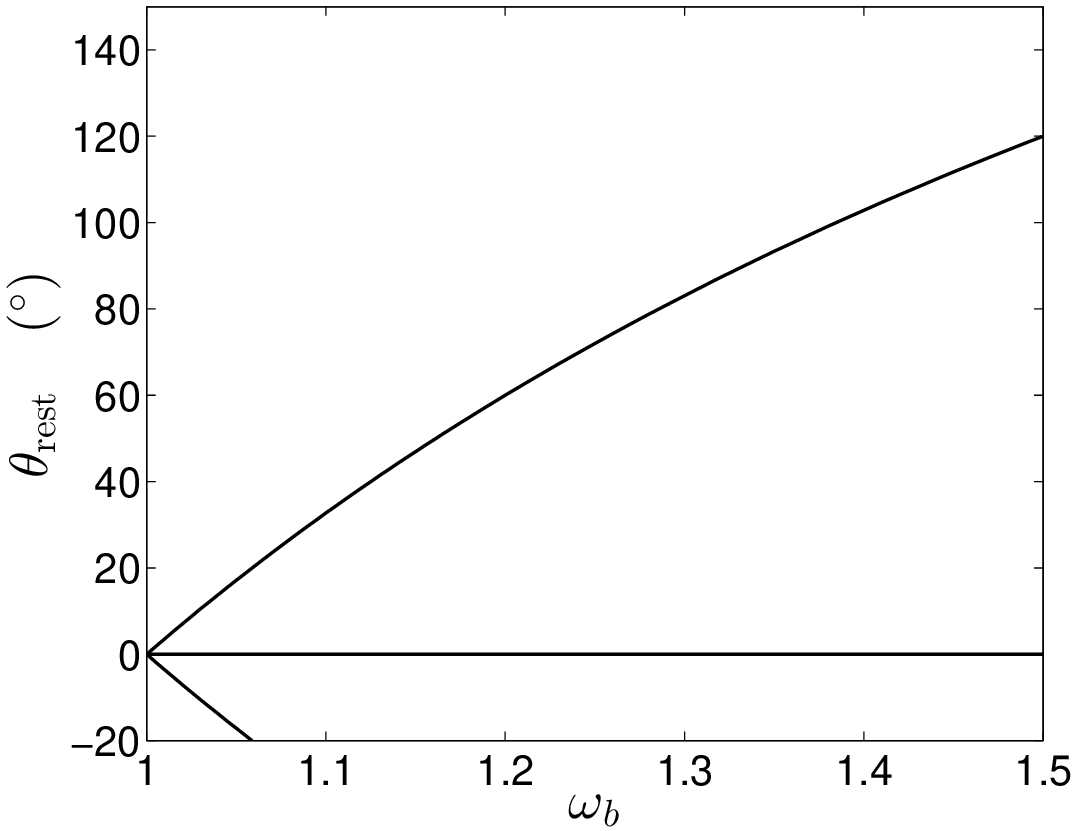}
\end{center}
\caption{ System with quartic hard potential (QH): ({\bf Left:}) Value of the initial coordinate $u_0$ for a single oscillator with hard potential as a function of the frequency $\omega_b$. ({\bf Right:}) Angles of the Floquet eigenvalues with zero coupling corresponding to the $N-2$ oscillators at rest and the single excited oscillator.
 }
\label{fig_qhsinglewb}
\end{figure}

\subsection{Breathers with hard quartic potential}
When we increase the coupling parameter $\kappa$ starting with a single excited oscillator, by path continuation, we can obtain the Fourier components of $z_{k,n}$ of the time-reversible, single breather with frequency $\omega_b$, given by $u_n(t)=z_{0,n}+\sum_{k=1}^{k_m}z_{k,n}\cos(k\omega_b t )$, the Floquet eigenvalues with modulus corresponding to the phonons having frequencies between $\omega_0$ and $\omega_\textrm{max}=\sqrt{\omega_0^2+4\kappa}$. Therefore, the maximum value for $\kappa$ corresponds to $\omega_{max}=\omega_b$ or $\kappa_\mathrm{max}=(\omega_b^2-\omega_0^2)/4$. In Fig.\,\ref{fig_brh12s1profile}-left we can observe the profile of breathers when increasing $\kappa$. For example, for $\omega_b=1.2$, $\kappa_\mathrm{max}=0.11$ and $0.31$ for $\omega_b=1.5$. Figure\,\ref{fig_brh12s1profile}-right shows the dependence of the maximum value of the initial coordinates of the breather and its energy as functions of $\kappa$. As predicted, the breather cannot be continued from $\kappa\simeq 0.11$.

We will use the value of $\kappa=0.05$ so that the coupling is significant but breathers are well localized and not too close to the instability.

\subsection{System with soft quartic potential}
\label{subsec:app:qs}
We can modify the system so that the on-site potential becomes soft changing the sign in front of the anharmonic term. That is, $U(u_n)=\omega_0^2(\frac{1}{2}u_n^2-\frac{1}{4}|s|\,u_n^4)$. In this case, the potential has a maximum at $u_n=1/\sqrt{|s|}$. If $u_n$ crosses the potential barrier, then there is an infinite negative potential well, which is not physically sound. Keeping with an interparticle distance of unity, it is therefore convenient to use $|s|=1$, so that the values of $|u_n|<1$ are outside the negative well and inside the safe well.

\begin{figure}[t]
\begin{center}
\includegraphics[width=0.49\textwidth]{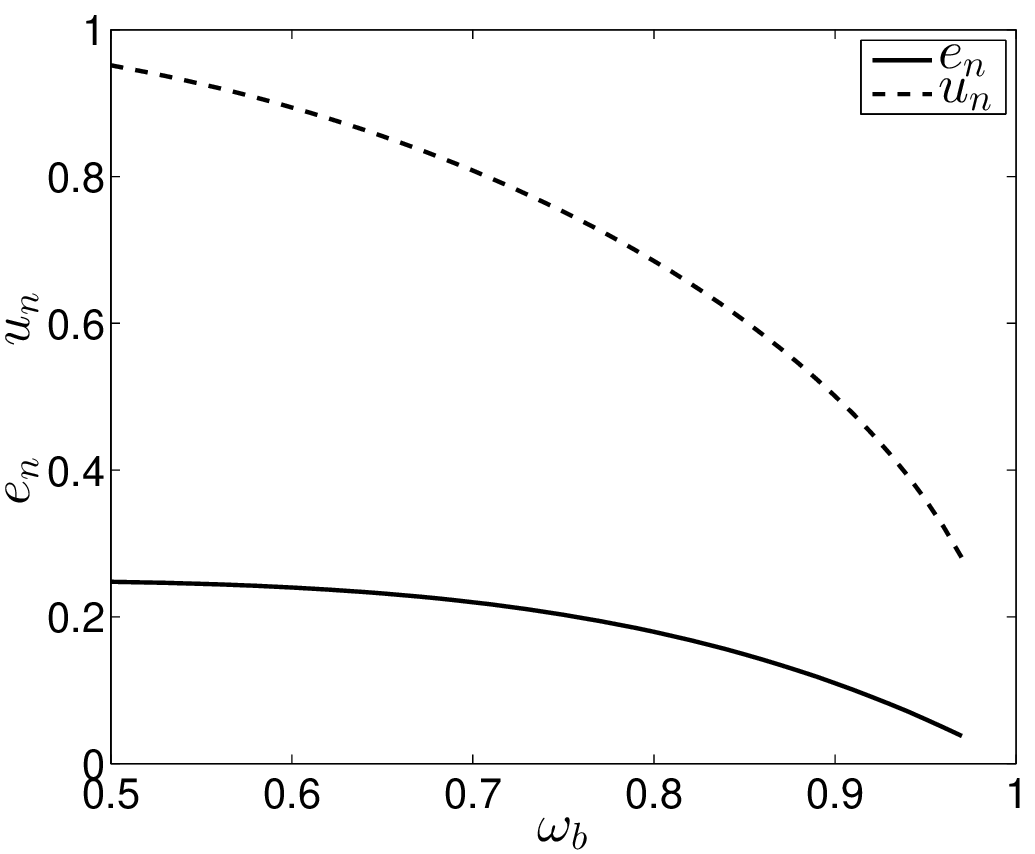}\,\,
\includegraphics[width=0.49\textwidth]{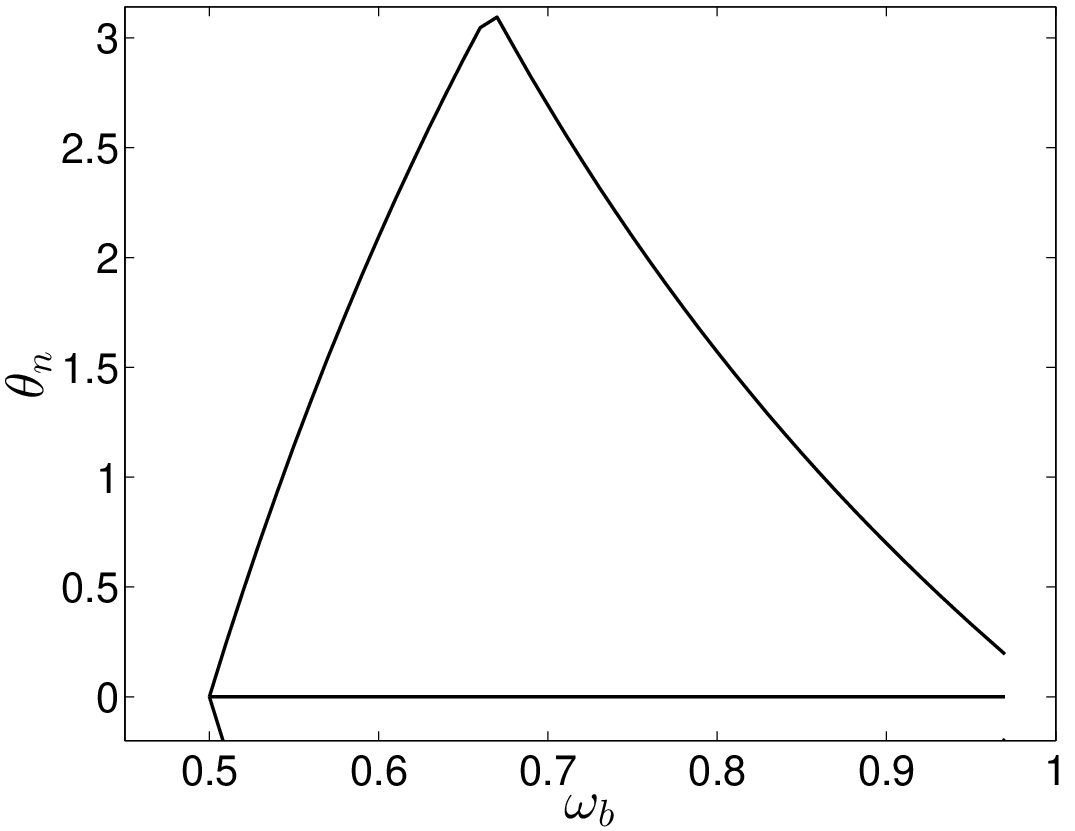}
\end{center}
\caption{ ({\bf Left:}) Value of the initial coordinate $u_0$ for a single oscillator with soft potential with
$|s|=1$ as a function of the frequency $\omega_b$. ({\bf Right:}) Angles of the Floquet eigenvalues with zero coupling corresponding to the $N-2$ oscillators at rest and the single excited oscillator for the same system.
 }
\label{fig_singlesoft}
\end{figure}

In this case, the single oscillator has a frequency that becomes smaller as the amplitude increases and is, therefore, smaller than $\omega_0$. The Floquet eigenvalues corresponding to the oscillators at rest are $\exp(\ii 2\pi \omega_0/\omega_b)$ and have angles larger than $2\pi$ that will attain $3\pi$ at $\omega_b=2/3\omega_0$ and $4\pi$ at $\omega_b=4\pi$. At those frequencies, the rest of the eigenvalues will cross with the possibility of leaving the unit circle, and the solution will become unstable for the coupled system. Both the amplitudes and the Floquet angles are shown in Fig.\,\ref{fig_singlesoft}.

\begin{figure}[ht]
\begin{center}
\includegraphics[width=0.49\textwidth]{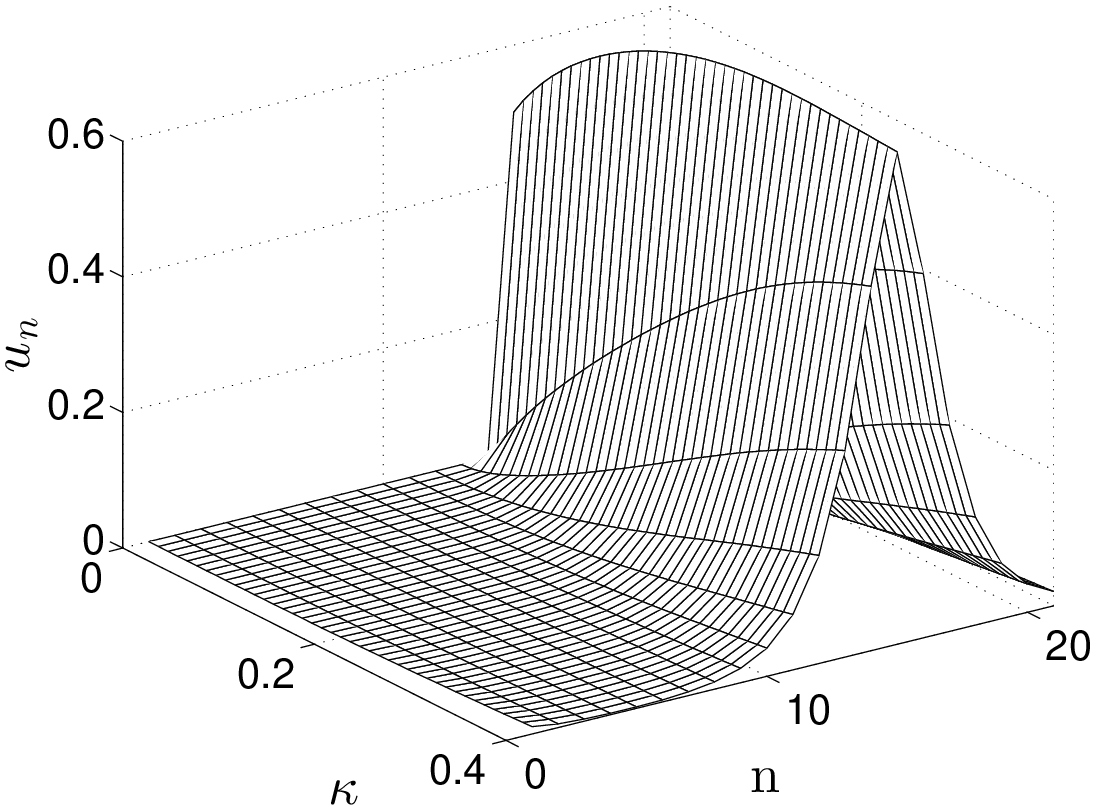}\,\,
\includegraphics[width=0.49\textwidth]{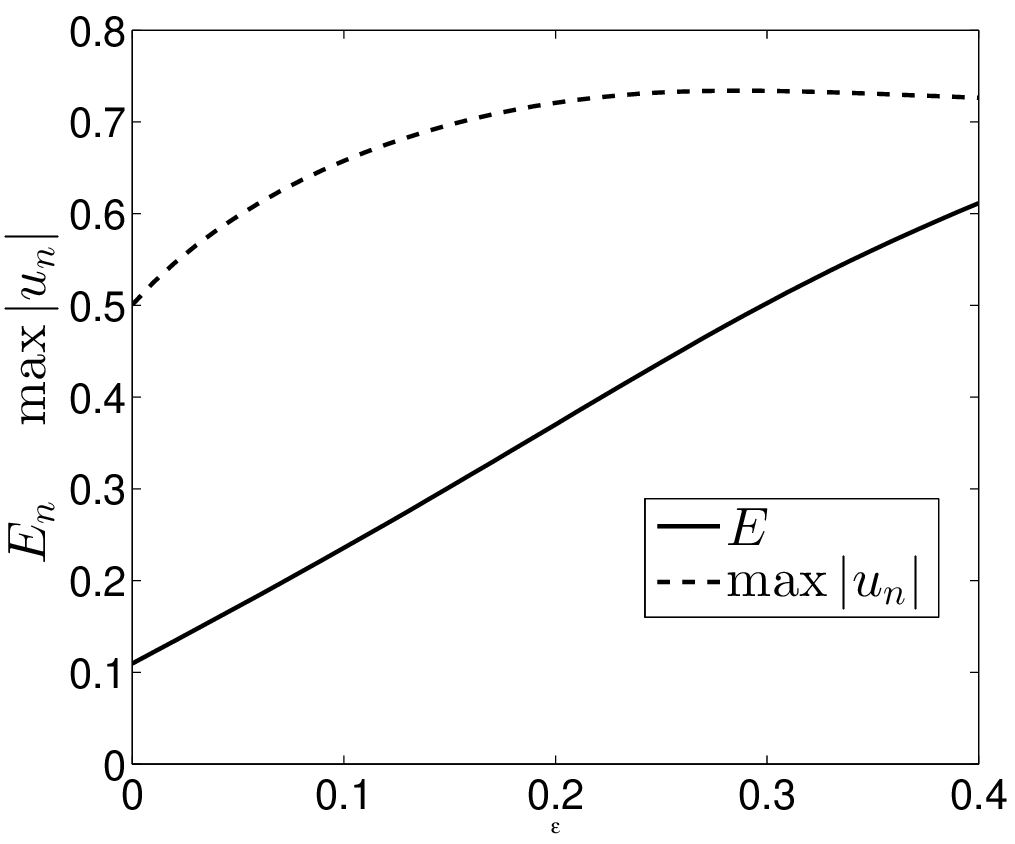}
\end{center}
\caption{ System with quartic soft potential (QS): ({\bf Left:}) Profile of breathers with soft potential and $\omega_b=0.9$ as a function of the coupling parameter $\kappa$. For $\kappa=1.1$, the breather becomes unstable. ({\bf Right:}) Maximum value of the initial condition (dashed line) and energy (continuous line) as functions of the coupling parameter $\kappa$.
 }
\label{fig_brs09profile}
\end{figure}

We can obtain the breathers by path continuation for $\kappa >0$. The breathers are bell-shaped as they derive from the mode with zero wavenumber and therefore the oscillators vibrate in phase. As the phonon maximum frequency increases to $\omega_\mathrm{max}=\sqrt{\omega_0^2+4\kappa}$, it will eventually coincide with the second harmonic of the breather, and the path continuation will fail. This corresponds to $\omega_b=0.9$ and $\kappa=0.49$, a fairly large value that we do not consider, limiting $\kappa$ to 0.4 in this calculation. The profile, energies, and maximum amplitude are shown in Fig.\,\ref{fig_brs09profile}.

\section{The Frenkel-Kontorova, Lennard-Jones system (FKLJ)}
\label{sec:app:fklj}
\setcounter{figure}{0}
The Hamiltonian for the FKLJ system is given in Eq.\,\eqref{eq:hamiltonianfklj}. We choose the parameters so that it is easy to compare with the previous systems (QS). Specifically, the on-site potential and the interatomic potential have the same first (zero) and second derivatives, i.e., $\omega_0=1$, and, as the negative quartic potential has a potential barrier at $x=\pm 1$, we also impose that condition. That means that the lattice distance is $\sigma=2$. The third condition is that the relative coupling parameter at low values of $u_n$ is also equal to the quartic soft potential coupling parameter.

We choose a formulation (i.e., \eqref{eq:hamiltonianfklj}) so that the scaling becomes obvious:
\begin{align}
\begin{split}
H=\sum_n \Biggl(\frac{1}{2}p_n^2 + & U_0\left(1-\cos\left(2\pi\frac{u_n}{\sigma}\right)\right) \\
&+V_0\left[1+\frac{1}{\left(1+\fracc{u_{n+1}-u_n}{\sigma}\right)^{12}}-\frac{2}{\left(1+\fracc{u_{n+1}-u_n}{\sigma}\right)^6}\right]\Biggl)\,,
\end{split}
\label{eq:app:hamiltonianfklj}
\end{align}
where $U_0=\omega_0^2 {\sigma^2}/{(2\pi)^2}$, and $\sigma$ is both the interatomic distance and the distance of the minimum of the LJ potential, i.e., the separation at equilibrium between atoms. $V_0$ is the depth of the LJ potential, which is shifted so that the minimum energy is zero. The harmonic FK frequency is $\omega_0$ as shown below.
The Taylor series up to the second power of the Hamiltonian \eqref{eq:app:hamiltonianfklj} yields:
\begin{equation}
H_L=\sum_n \left(\frac{1}{2}p_n^2+ \frac{1}{2}\omega_0^2 u_n^2 +
\frac{1}{2}\omega_0^2\kappa (u_{n+1}-u_{n})^2\right)\,,
\label{eq:hamiltonianfkljlinear}
\end{equation}
where $\kappa$ is the relative coupling with respect to the on-site potential and it is given by $\kappa=72 V_0/\omega_0^2\sigma^2$.
In the BEL system\,\cite{bajars-physicad2015},  $\omega_0=2\pi$ and $\sigma=1$, so $\kappa=72 V_0/(2\pi)^2$. In the present work, as $\omega_0=1$ and $\sigma=2$,  $\kappa=72 V_0/2^2=18 V_0$, or $V_0=2^2\kappa/72=\kappa/18$, that is, $V_0=0.00278$ for $\kappa=0.05$ and $V_0=0.00556$ for $\kappa=0.1$.

\begin{figure}[t]                                                                                                                                    \begin{center}
\includegraphics[width=0.49\textwidth]{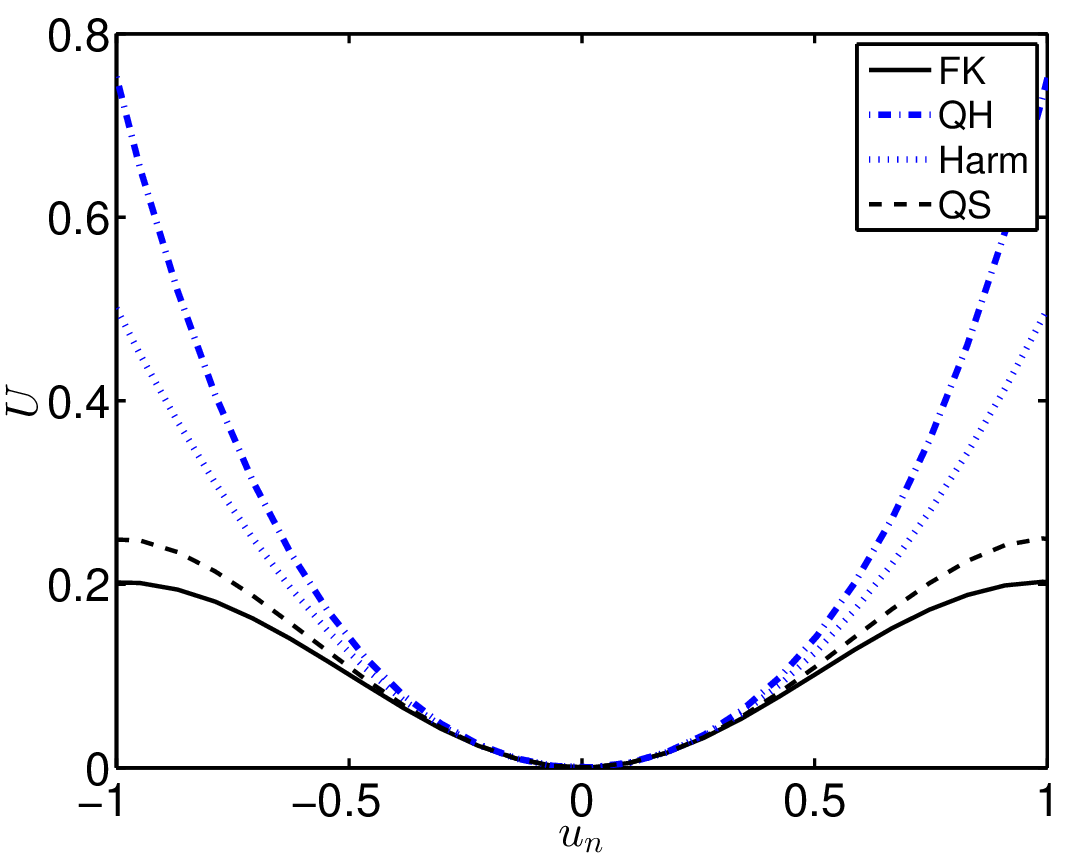}\,\,
\includegraphics[width=0.49\textwidth]{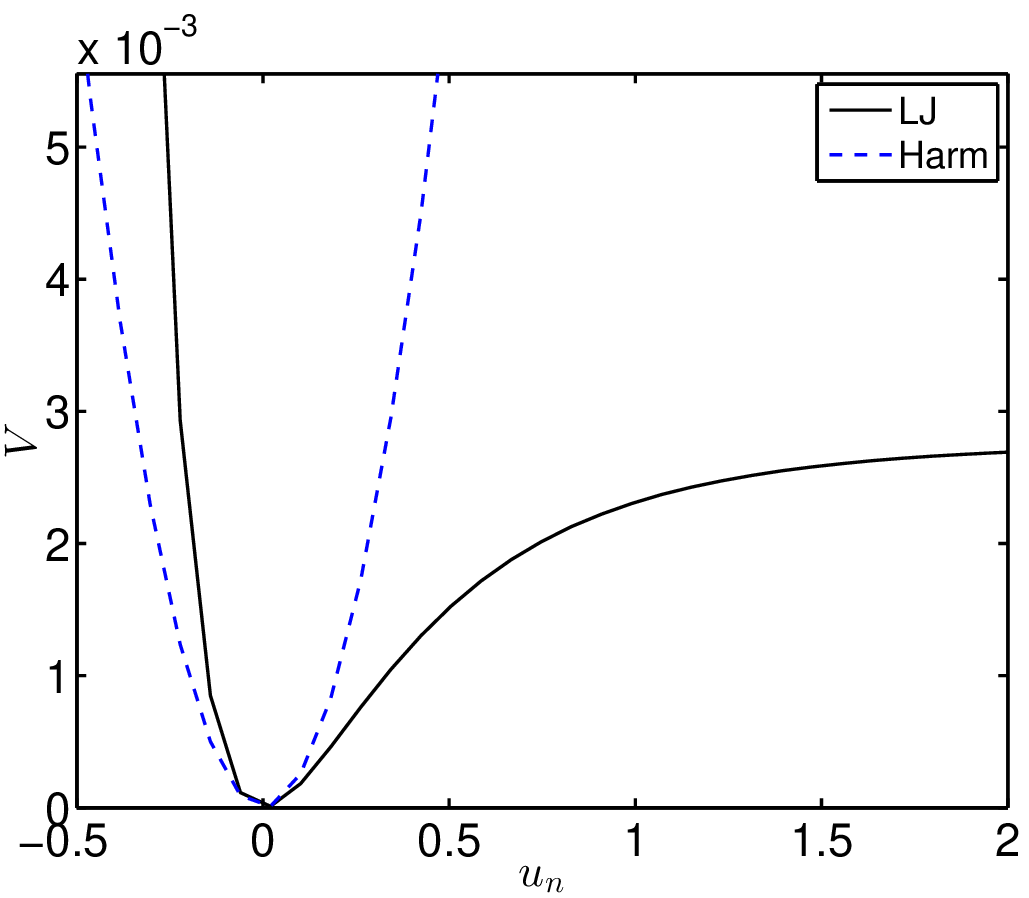}
\end{center}
\caption{ ({\bf Left:}) Plot of the different on-site potentials $\kappa=0.05$.
({\bf Right:}) Plot of the different interatomic interaction potentials for $\kappa=0.05$.
 }
\label{fig_compUV}
\end{figure}

The comparison of the different potentials is shown in Fig.\,\ref{fig_compUV}-left. We can see that the FK potential well has small energy, being the repulsive interaction the most important component.

From the anticontinuous limit, we can construct bell-shaped, single breathers with frequencies $\omega_b$ below the bottom of the phonon band $\omega_0=1$, until $1.5\omega_b$ hits the top of the phonon band $\omega_t=\sqrt{\omega_0^2+4\kappa}$.

\begin{figure}[htb]                                                                            \begin{center}
\includegraphics[width=0.49\textwidth]{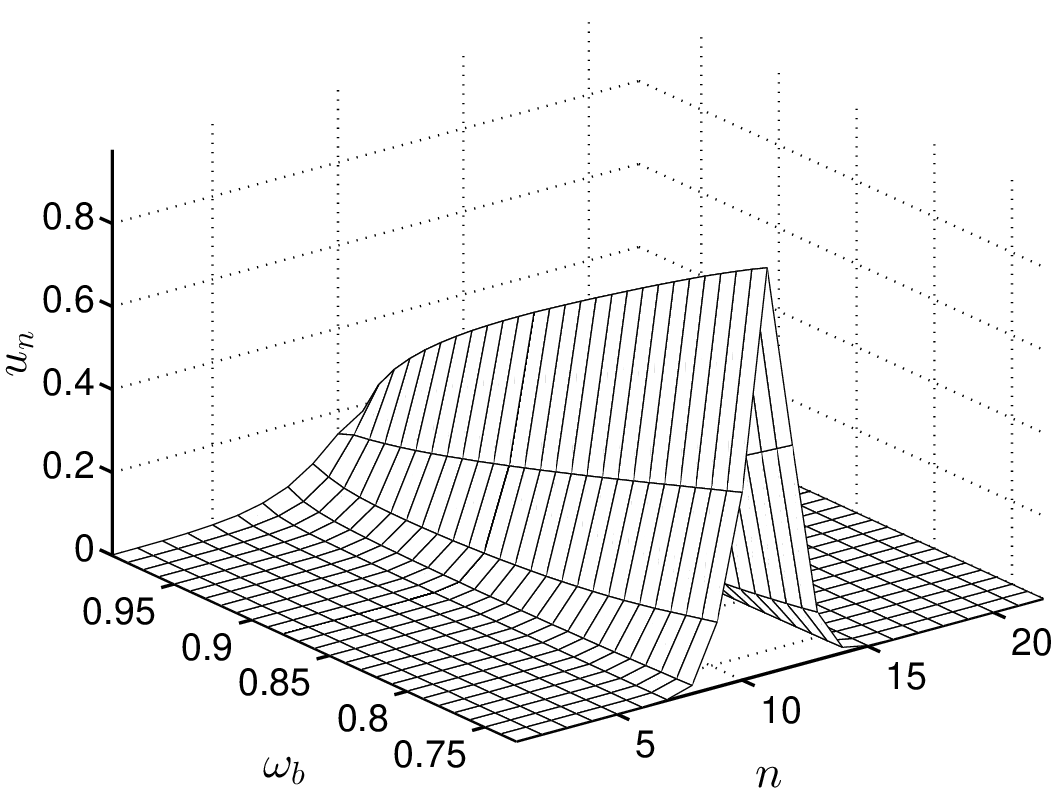}\,\,
\includegraphics[width=0.49\textwidth]{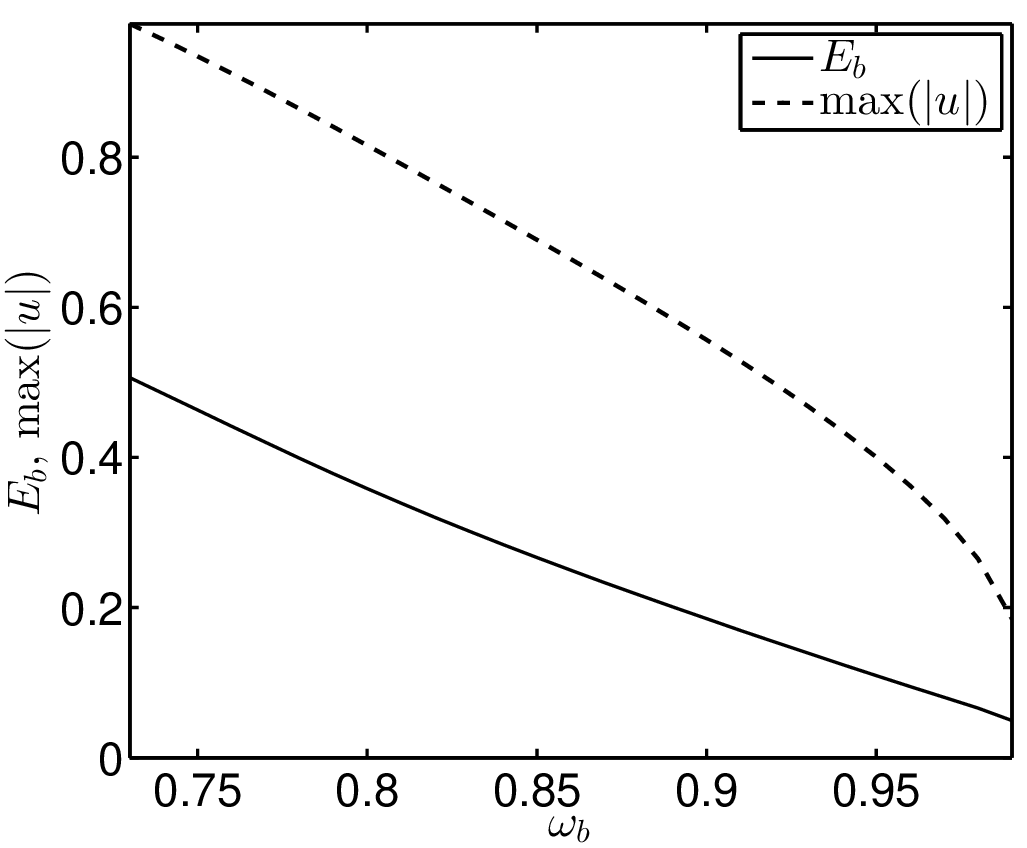}
\end{center}
\caption{ System FKLJ: ({\bf Left:}) Profile of the breathers  with $\kappa=0.05$ as a function of the frequency.
({\bf Right:}) Plot of amplitude and energy with $\kappa=0.05$ as a function of the frequency.
 }
\label{fig_fkljbreather}
\end{figure}

\section{Breathers in the Josephson junction network (JJN)}
\label{sec:app:jjn}
\setcounter{figure}{0}
As presented in Sec.\,\ref{sec:jjn}, the JJN system \eqref{eq:Hjjn} can also be written as:
\begin{equation}
H=\sum_n \left(\frac{1}{2}p_n^2+\kappa (V(q_{n+1}-q_n)-V(q_n-q_{n-1}))\right)\,,   
\label{eq:app:Hjjn}
\end{equation}
with $V(x)=1-\cos(x)$.

This system has no on-site potential and it is a non-polynomial variant of the Fermi-Pasta-Ulam system (FPU)\,\cite{ford_FPU1992}, for which breathers are studied in Ref.\,\cite{sanchez-rey2024}, that provides conditions for the existence of breathers that we use below.

The corresponding  dynamical equations $\dot p_n=\ddot q_n=-\partial H/\partial q_n$ are:
\begin{equation}
\ddot q_n= \kappa\left(V'(q_{n+1}-q_n)-V'(q_n-q_{n-1})\right). 
\label{eq:app:dynamicjjn}
\end{equation}

To use the results in Ref.\,\cite{sanchez-rey2024}, we have to re-scale the time so that $\kappa$ is substituted by the unity. It is easy to see that $\kappa=c^2$, with $c$, the sound velocity, that is, the limit of both the phase velocity and group velocity when the wavenumber $k\rightarrow 0$.  Then,  $t=\tilde t/c$, $\omega=c\tilde \omega$, where the variables with tilde are the ones in the cited reference.

Then, we need to calculate the four derivatives of $V$ at zero:
\begin{equation}
V'(0)=0;\quad V''(0)=1;\quad K_3\equiv V^{(3)}(0)=0;\quad K_4\equiv V^{(4)}(0)=-1\,.
\label{eq:fourderivatives}
\end{equation}

The phonon band is given by $\omega(k)=2 c\sin(k/2)$.  The minimal value is $\omega_\text{min}=\omega(0)=0$, i.e., it is an acoustic dispersion law, and the maximum is $\omega_\text{max}=\omega(\pi)=2 c $. The existence of small amplitude breathers with frequency above $\omega_\text{max}=\omega(\pi)=2 c$ depends on the sign of the quantity $B=0.5 K_4-K_3^2=0.5 (-1)-0^3=-0.5$. If it is negative, as in our case, then there are no small amplitude breathers.

With $B<0$, there are large amplitude breathers if $K_4>0$ and $|K_3|<\sqrt{3K_4}$, conditions that do not hold in our system. Therefore, we are not in the condition where there are proofs of breather existence.

To find out which type of localization is there we analyze the evolution of the system for different values of the parameters, for $E_\text{b}=0.40$ and $\kappa=0.05$. The localization shows a relatively long thermalization time as shown in Sec.\,\ref{sec:jjn}.

\end{document}